\documentclass[galaxies,article,accept,moreauthors]{Definitions/mdpi}

\firstpage{1} 
\pubvolume{1}
\issuenum{1}
\articlenumber{0}
\pubyear{2026}
\copyrightyear{2026}
\externaleditor{Firstname Lastname} 
\datereceived{25 September 2026} 
\daterevised{28 October 2026} 
\dateaccepted{31 October 2026} 
\datepublished{ } 
\usepackage{changepage}
\definecolor{sb}{RGB}{70,130,180}
\definecolor{dg}{RGB}{77,140,80}
\definecolor{rb}{RGB}{65, 105, 225}
\definecolor{dy}{RGB}{0, 0, 0}
\definecolor{gs}{RGB}{210,140,70}
\definecolor{vk}{RGB}{128,0,128}

\usepackage[russian,english]{babel}
\usepackage{textcomp}
\usepackage{siunitx}
\usepackage{amsmath}
\usepackage{xcolor}

\newsavebox{\glassimage}

\Title{FAIR-Compliant Architecture for Heterogeneous Astronomical Data: KazVO Framework} 

\Author{Ildana Izmailova*\orcidA{},
Denis Yurin \orcidB{},
Yerlan Aimuratov *\orcidC{},
Maxim Makukov \orcidD{},
Aleksander Serebryanskiy \orcidE{},
\linebreak  Saule Shomshekova \orcidF{},
Vitaliy Kim \orcidG{},
Chingis Omarov \orcidH{},
Adel Umirbayeva \orcidI{},
Laura Aktay \orcidJ{},
Dana Kuvatova \orcidK{},
Anton Gluchshenko \orcidL{},
Gulnara Suliyeva \orcidM{},
Maxim Krugov \orcidN{},
Nadezhda Vaidman \orcidO{},
Daulet Anarbek \orcidP{},
\linebreak  Inna Reva \orcidQ{},
Gauhar Aimanova \orcidR{},
Rashit Valiullin \orcidS{}
and Raushan Kokumbaeva \orcidT{}
}

\AuthorNames{Ildana Izmailova, Denis Yurin, Yerlan Aimuratov, Maxim Makukov, Aleksander Serebryanskiy, Saule Shomshekova, Vitaliy Kim, Chingis Omarov, Adel Umirbayeva, Laura Aktay, Dana Kuvatova, Anton Gluchshenko, Gulnara Suliyeva, Maxim Krugov, Nadezhda Vaidman, Daulet Anarbek, Inna Reva, Gauhar Aimanova, Rashit Valiullin and Raushan Kokumbaeva} 

\address[1]{Fesenkov Astrophysical Institute, Observatory 23, Medeu District, Almaty 050020, Kazakhstan}

\corres{\hangafter=1 \hangindent=1.05em \hspace{-0.82em} Correspondence: izmailova@fai.kz (I.I.); aimuratov@fai.kz (Y.A.)}

\abstract{This paper presents the architecture of the Kazakhstani National Virtual Observatory (KazVO)---an International Virtual Observatory Alliance (IVOA) compliant node that unifies the heterogeneous observational datasets of the Fesenkov Astrophysical Institute (FAI). The  architectural foundation of the system is built upon a partitioned Data Lake, a German Astrophysical Virtual Observatory (GAVO) Data Center Helper Suite (DaCHS) publishing backend, and {a PostgreSQL database in which metadata for published observational products are mapped to IVOA ObsCore, while catalogs and simulation grids use dedicated relational tables}. We follow Open Science principles by introducing a metadata-only showcase model based on a four-class data embargo mechanism. {Machine-to-Machine} services deployed via IVOA protocols {are exposed through the KazVO publishing registry registered in the global Registry of Registries (RofR)}, enabling  analysis of Kazakhstani observational assets within external clients such as Tool for OPerations on Catalogues And Tables (TOPCAT), Aladin, and PyVO. {
The current implementation combines operational VO publication services and targeted photometric processing with scheduled, batch-mode data ingestion, supporting access to published historical and modern observational holdings. Replication has been validated on pilot datasets, while routine deployment at the observing sites and automated reduction of modern spectroscopic streams remain under development. The architecture provides a framework for progressively integrating these components into a complete data lifecycle.}}


\begin{document}


\section{Introduction}

Modern astrophysics requires not only continuous accumulation of observational and theoretical data, but also provision of open, standardized, and equitable access to these resources. Aligned with the global Open Science paradigm \citep{unesco2021openscience}, the scientific productivity of modern observatories directly depends on their integration into unified digital ecosystems that combine automated data processing services, high-performance computing, and accessible large-scale archives \citep{reitze2024, zhang2015, aimuratov2025}. {Experience from major astronomical facilities highlights the importance of software infrastructure, processing pipelines, and accessible archives in supporting the scientific use and reuse of observational data~\citep{astro2020, norris2006, norris2007}. Neglecting these computational foundations significantly reduces the actual scientific output of observational facilities.}

This necessity stems from a fundamental paradigm shift in modern astronomy: a transition from model-driven verification (``hypothesis-to-observation'') to data-driven discovery (``data-to-hypothesis'') \citep{desouza2025}. Astronomy has evolved from passive viewing into an active, algorithmic querying of the sky, where observations form a basis for uncovering previously unknown phenomena. In this data-intensive era, modern astronomical datasets are defined by the ``8-V'' Big Data framework—encompassing Volume, Variety, Velocity, Veracity, Value, Visualization, Validity, and Variability. Extracting scientific value through advanced Statistical Analysis and Machine Learning (ML) techniques requires absolute data interoperability, rigorous quality control, and automated reproducibility across the entire data lifecycle \citep{huppenkothen2023, brescia2024}.

Within this evolving paradigm, astronomical archives have transformed from passive storage facilities into active research environments \citep{norris2007, erard2025sf2a}. Bringing highly heterogeneous observational collections to scientific maturity requires strict adherence to the FAIR principles (Findable, Accessible, Interoperable, Reusable) \citep{wilkinson2016fair, landais2024adass, otoole2024adass} and standards established by the IVOA \citep{dowler2021arch}. Standardized IVOA protocols---including Table Access Protocol \mbox{(TAP) \citep{dowler2019tap}}, Simple Image Access Protocol (SIAP) \citep{dowler2015sia}, and Simple Spectral Access Protocol (SSAP)~\mbox{\citep{tody2012ssap}---ensure} machine-readability and programmatic access required by modern ML analytical workflows \citep{huppenkothen2023}, while Digital Object Identifiers (DOIs) guarantee study reproducibility and citation tracking \citep{landais2024adass, landais2026ivoa}. The archival services of major facilities and data centers, including HST, ESO, and CDS, support scientific investigations that extend beyond the original observing programmes \citep{norris2006, astro2020, romaniello2023eso, lesteven2025lisa}.

Adopting this global framework enables national and regional research centers to mitigate the challenges of data isolation \citep{zhang2015}. Transitioning from legacy local archives to flexible, IVOA-compliant digital ecosystems ensures seamless handling of heterogeneous datasets while establishing regional observatories as fully fledged nodes within the global scientific network. Prominent implementations range from established national platforms, such as China-VO \citep{li2017}, Russian Virtual Observatory~(RVO) \citep{dluhznevskaya2018}, and Armenian Virtual Observatory (ArVO) \citep{mickaelian2025}, to the Integration-by-Design framework realized within the Astronomical Hub~(AstroHub) infrastructure \citep{aimuratov2025}.

For the FAI, addressing this global necessity has not required building a new ecosystem from scratch, but has rather relied on adopting established IVOA standards to homogenize local datasets and provide programmatic access. Driven by the accumulation of extensive historical archives and the rapid expansion of physical telescope infrastructure at the high-altitude Assy-Turgen Observatory~(ATO), {developing KazVO using IVOA standards and guided by the FAIR principles became a central objective}. This platform unifies highly heterogeneous national data streams: ranging from digitized historical photographic plates (1950--1997)---recognized as a vital baseline for long-term time-domain \mbox{astrophysics \citep{hudec2018, hudec2019, wenjing2007}---to} real-time {monitoring time series} targeting near-Earth space \citep{rovetto2016} and deep-space objects \citep{nebot2019}.

In this paper, we present the architecture and technical implementation of the KazVO, focusing on the pipelines and methodologies applied to {improve the findability, accessibility, interoperability, and reusability of heterogeneous regional datasets}. The remainder of this paper is structured as follows: Section~\ref{sec:materials_methods} details the observational collections, optical facilities, computing environment, and the {multi-stage FAIR-ification workflow}, including metadata extraction, IVOA protocols, and remote ingestion schemes. Section~\ref{sec:results} demonstrates the operational KazVO architecture, automated processing workflows, TAP service validations, and embargo mechanisms. {Section~\ref{sec:disc} discusses our architectural choices in the context of regional virtual observatory development}, and Section~\ref{sec:concl} summarizes key conclusions.


\section{Materials and Methods} \label{sec:materials_methods}

\begin{figure}[H]
\includegraphics[width=0.9\textwidth]{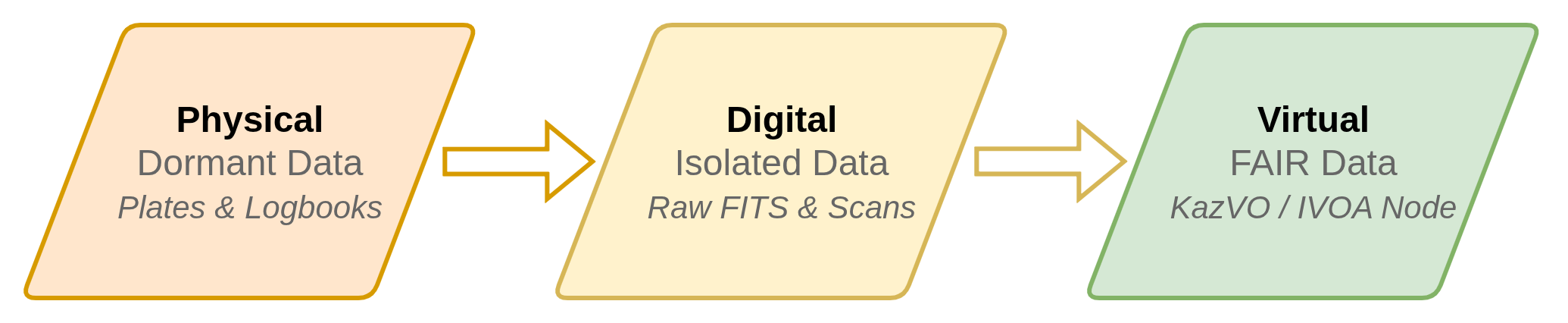}
\caption{The workflow of astronomical data evolution in KazVO: transitioning from non-FAIR physical heritage assets and isolated raw digital streams to a fully interoperable IVOA-compliant VO node.}
\label{fig:data_evolution}
\end{figure}

The overall methodological framework of KazVO relies on a data evolution model that ensures the sequential transformation of astronomical resources from analog and isolated local digital states into full interoperability within the VO ecosystem (Figure~\ref{fig:data_evolution}).

This section is structured as follows: Section~\ref{subsec:data_materials} describes the observational materials (physical heritage assets and digital Flexible Image Transport System (FITS) streams); Section~\ref{subsec:infrastructure} outlines the instrumental and computing environment used for data acquisition and storage; and Section~\ref{subsec:fair_pipeline} details the pipeline for data standardization and integration according to FAIR and IVOA standards.

\subsection{Heterogeneous Data Collections: Observational Materials}\label{subsec:data_materials}

The observational basis of this work consists of astronomical data accumulated at FAI over more than 70~years, with structural heterogeneity posing the primary challenge. The dataset encompasses two main categories of materials: physical analog media, glass photographic plates and films from historical archives, and incoming digital data streams from modern detectors. Integrating these heterogeneous formats into a unified machine-readable ecosystem requires applying the FAIR principles to ensure their standardization and global accessibility.

\subsubsection{Physical Heritage Observational Materials}

\begin{figure}[H]
    \centering
    \begin{minipage}[c]{0.55\textwidth}
    \centering
   \hspace*{-31mm}     \includegraphics[width=\linewidth]{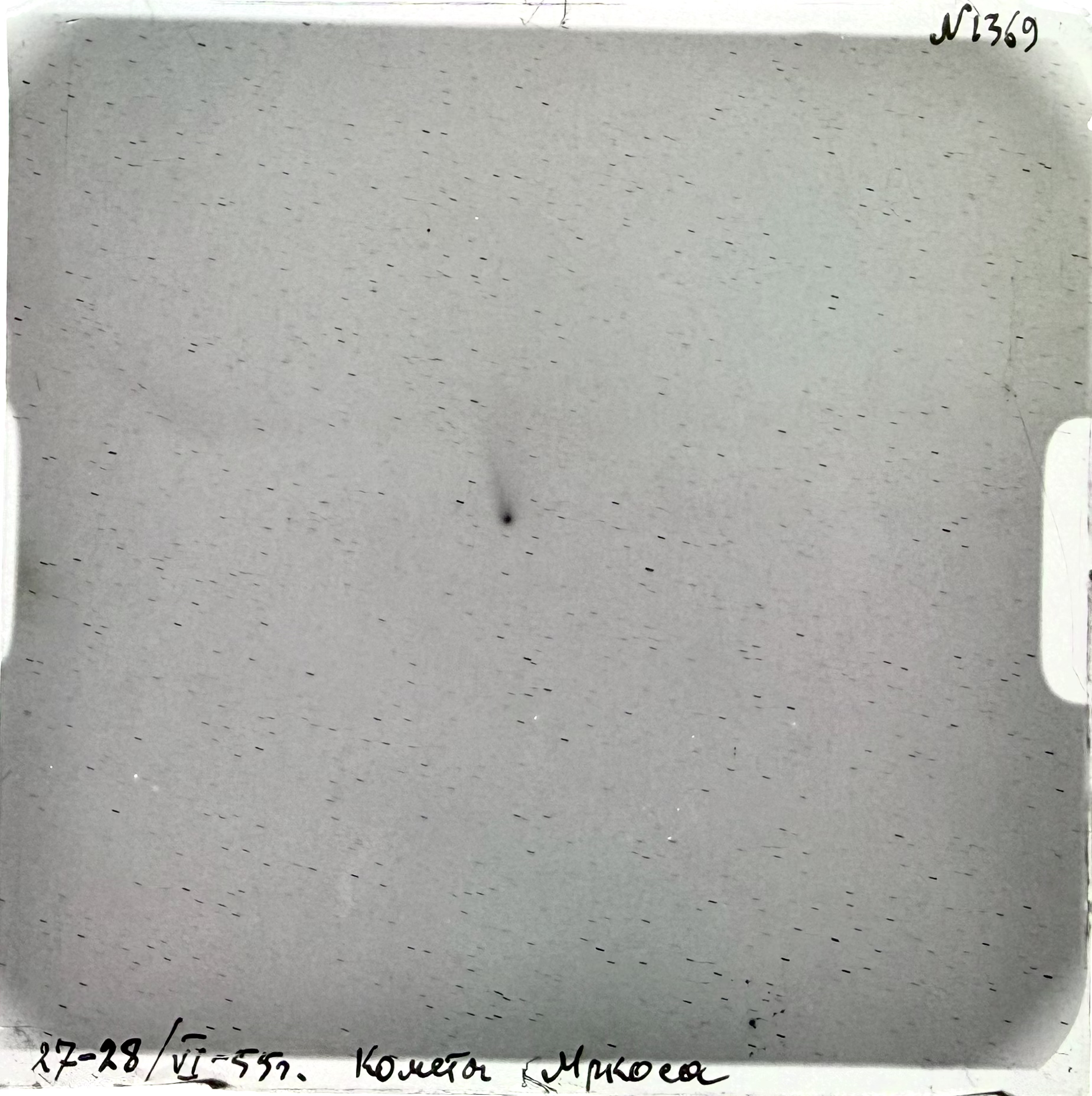}
        \par\smallskip
     \hspace*{-31mm}      (\textbf{a})
    \end{minipage}

    \vspace{0.4cm}
    
    \begin{minipage}[c]{0.75\textwidth}
        \centering
     \hspace*{-31mm}    \includegraphics[width=\linewidth]{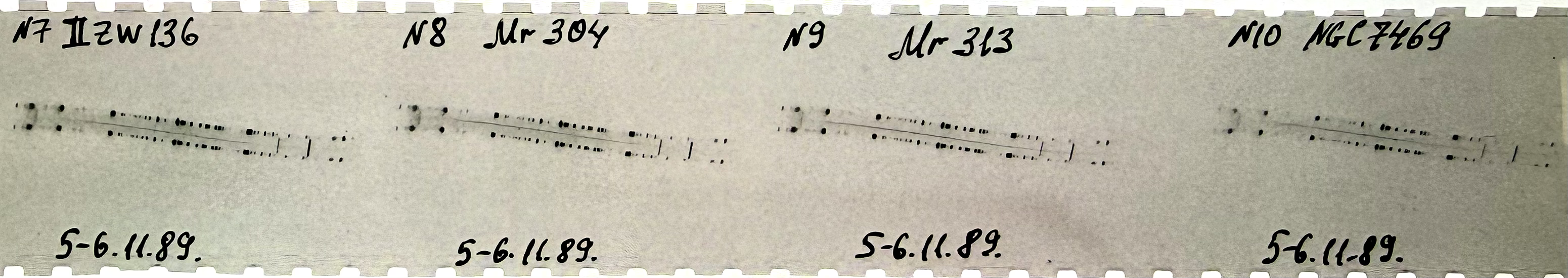}
        \par\smallskip
   \hspace*{-31mm}      (\textbf{b})
    \end{minipage}
    
    \caption{Representative physical heritage assets from the FAI archival collection: (\textbf{a})~the glass plate of Comet Mrkos (C/1955~L1) acquired at the Maksutov meniscus telescope (ASI-2) on 27~June~1955. Plate bottom handwritten caption denotes the frame acquisition date and object name ``27--28/VI-55. Comet Mrkos'';} (\textbf{b})~a film strip containing spectra recorded on photographic film obtained at the AZT-8 telescope on 25~November~1989.
    \label{fig:films_exmpl}
\end{figure}

The historical archive of FAI photographic plates and films, informally referred to as the ``Glass Library'', spans the period from 1950 to 1997 and serves as the primary physical material for this study. The collection comprises approximately 20,000~analog negatives, including 7700~glass photographic plates and 12,000~photographic films (representative samples are shown in Figure~\ref{fig:films_exmpl}). Observations were conducted using three main instruments: the 50-cm Maksutov meniscus telescope (ASI-2), the 39.7-cm Schmidt camera, and the 70-cm AZT-8 reflector, see Table~\ref{tab:archive_telescopes}. The quantitative distribution of digitized legacy products integrated into KazVO is presented in Section~\ref{subsec:res_holdings}.

\begin{table}[H]
\caption{Optical specifications and physical carrier types of FAI historical instruments (1950--1997). The table details the optical parameters and physical carriers of historical facilities. Total volumes, file counts, and FAIR data levels of digitized products integrated into KazVO are presented in Section~\ref{subsec:res_holdings}.} 

\label{tab:archive_telescopes}

\begin{adjustwidth}{-\extralength}{0cm}
\footnotesize
\begin{tabular*}{\fulllength}{@{\extracolsep{\fill}}lccccccccccc@{}}
\toprule
\textbf{Telescope \textsuperscript{1}} & \textbf{D [mm]} & \textbf{f [mm]} & \textbf{Carrier Type} & \textbf{Frame Size [mm]}
& \textbf{FoV/Slit} & \textbf{Period [yr]} & \textbf{Observation Type} \\
\midrule

Schmidt & 397 & 773 & Glass Plate & $90 \times 120$ & $6.2^\circ \times 8.4^\circ$
& 1964--1989 & Direct Photometry \\

ASI-2 & 500 & 1200 & Glass Plate & $90 \times 98$
& {$4.3^\circ \times 4.7^\circ$} & 1950--1997 & Direct Photometry \\

AZT-8 & 700 & 11,200/28,000 & Photographic Film & $180 \times 20$ & $0.05'$--$0.17'$
& {1970--1997} & Slit Spectroscopy \\

\bottomrule
\end{tabular*}


\end{adjustwidth}
\noindent{\footnotesize{\textsuperscript{1}~
\textbf{Schmidt}: Schmidt Camera (Catadioptric Astrophotographic Telescope), Carl Zeiss JENA, Jena, German Democratic Republic (GDR / East Germany); 
\textbf{ASI-2}: ``Astronomicheskiy Spektral'nyy Teleskop'' model No.~2, also known as Maksutov Meniscus Telescope, Gosudarstvennyy Opticheskiy Institut (Vavilov State Optical Institute), Leningrad, USSR;
\textbf{AZT-8}: ``Astronomicheskiy Zerkal'nyy Teleskop'' model No.~8, LOMO, Leningrad, USSR.}}
\end{table}

The observational materials combine survey and targeted photometry, as well as slit spectroscopy. The photometric archive includes survey fields, targeted stellar fields, and monitoring data of Solar System small bodies, specifically comets and asteroids. The spectroscopic dataset, acquired with the AZT-8 telescope using an FAI-developed focal spectrograph \citep{denisyuk2003}.
 and an UM-92 image intensifier tube (IIT), predominantly covers the optical wavelength range ($3700$--$8600$~\text{\AA}), forming a unique historical baseline for analyzing the spectral variability and kinematics of various classes of astrophysical objects.

\begin{figure}[H]
    \includegraphics[width=0.6\linewidth]{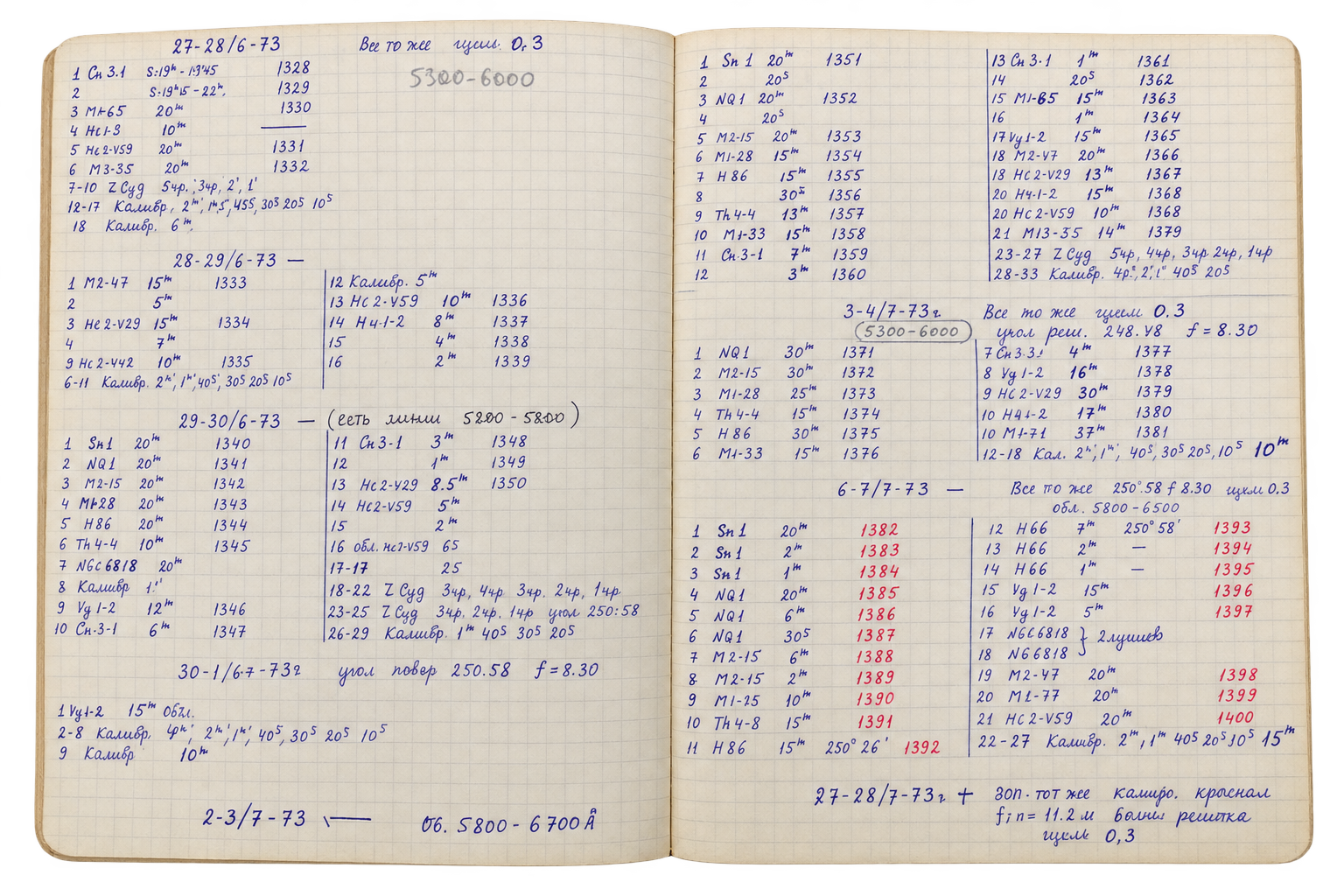}
    \caption{Representative pages from the historical FAI handwritten observational logbook. The image shows an unedited scan of historical handwritten logbooks (1950--1997); numerical formatting and punctuation within the handwritten text reflect original archival records related to dates, objects, observation time, observational setup, weather conditions, etc. The non-English terms (handwritten notes in Russian) represent the primary historical provenance and metadata of the astronomical logbook.}
    \label{fig:obslog}
\end{figure}

Primary metadata, such as coordinates, exposure times, grating characteristics, and observational conditions, have been preserved in dozens of volumes of original handwritten logbooks (Figure~\ref{fig:obslog}).

Both the analog negatives and the handwritten logbooks require translation into modern machine-readable standards. Integrating these resources into KazVO opens up key opportunities for time-domain astrophysical research.

\subsubsection{Digital Observational Data Material}

The second category of observational material consists of modern digital data streams regularly arriving from active instruments at the ATO and the Tien-Shan Astronomical Observatory (TShAO). Although these digital data contain informative metadata in their headers, prior to the deployment of KazVO, the nomenclature of their FITS keywords required unification to meet international IVOA standards, and astrometric reduction was performed only for selected observational sets. {Within KazVO, metadata for published observational products are mapped to the IVOA ObsCore data model \citep{ivoa.obscore2017}. Calibration levels describe the processing applied to individual products rather than their validation status: instrumental data in standard FITS format correspond to Level~1, calibrated products to Level~2, and enhanced products such as aligned, co-added science images to Level~3. The published collections are summarized in Section~\ref{sec:results}.}

\subsection{Observational Infrastructure and Computing Environment} \label{subsec:infrastructure}

The hardware foundation of KazVO relies on the FAI infrastructure, which integrates a distributed network of optical telescopes with a central computing cluster.

\subsubsection{Optical Facilities as Primary Data Generators}

Modern digital data streams in KazVO originate primarily from the optical facilities at ATO and TShAO. High-speed data transmission from ATO is provided via Starlink satellite terminals, whereas TShAO is connected via dedicated terrestrial links. Table~\ref{tab:telescopes} summarizes the specifications of active FAI telescopes whose data are systematically ingested into the central KazVO repository.

\begin{table}[H]
\caption{Optical specification and operational data generation rates of active FAI instruments integrated into KazVO. Data rate represents the average raw data throughput generated per active observing night. Total accumulated file counts, consolidated volumes, and FAIR data levels published within the KazVO repository are detailed in Section~\ref{subsec:res_holdings}. 
 Partner instruments within the AstroHub ecosystem \citep{aimuratov2025} are omitted here.}
\label{tab:telescopes}

\begin{adjustwidth}{-\extralength}{0cm}
\footnotesize
\begin{tabular*}{\fulllength}{@{\extracolsep{\fill}}lcccccccccc@{}}
\toprule
\textbf{Telescope \textsuperscript{1}} & \textbf{Obs.} & \textbf{D} & \textbf{f} & \textbf{Obs.} & \textbf{Detector} & \textbf{FoV / Slit} & \textbf{Commis.} & \textbf{Data Rate}  \\
{} & {} & \textbf{[mm]} & \textbf{[mm]} & \textbf{Type} & {} & {} & \textbf{[yr]}  & \textbf{[GB/night]}  \\
\midrule

AZT-20 & ATO & 1560 & 5720 & Spec/Phot & EMCCD/CMOS & $14'\times14'$/$3''$--$9''$ & 2017 & 3.97 \\
WFOS-70 & ATO & 660 & 900 & Photometry & CMOS QHY-6060 & $3.8^\circ\times3.8^\circ$ & 2025 & 11.27 \\
RC-500 & ATO & 508 & 1400 & Photometry & CMOS QHY-600 & $1.3^\circ\times1.0^\circ$ & 2019 & 16.59 \\
WFOS-40 & ATO & 400 & 550 & Photometry & CMOS QHY-600 & $3.8^\circ\times2.5^\circ$ & 2024 & 46.07 \\
Zeiss-1000 (E) & TShAO & 1016 & 6500 & Photometry & Apogee U9000 & $20'\times20'$ & 2013 & 2.71 \\
Zeiss-1000 (W) & TShAO & 1016 & 13,280 & Spectroscopy & Apogee CG42UV & $7.2'\times7.2'$ & 2013 & 1.09 \\

\bottomrule
\end{tabular*}

\vspace{2mm}

\end{adjustwidth}
\noindent{\footnotesize{\textsuperscript{1}~
\textbf{AZT-20}: ``Astronomicheskiy Zerkal'nyy Teleskop'' model No.~20, LOMO, Leningrad, USSR; 
\textbf{WFOS-70}: 70-cm aperture Wide-Field Optical System, AstroTechPribor, Almaty, Kazakhstan; 
\textbf{RC-500}: 500-mm aperture Ritchey-Chr\'etien reflective telescope, AstroSib, Novosibirsk, Russia; 
\textbf{WFOS-40}: 40-cm aperture Wide-Field Optical System, AstroTechPribor, Almaty, Kazakhstan; 
\textbf{Zeiss-1000 (E)}: ``East'' 1-m aperture Ritchey-Chr\'etien reflective telescope, Carl Zeiss JENA, Jena, German Democratic Republic (GDR / East Germany); 
\textbf{Zeiss-1000 (W)}: ``West'' 1-m aperture Ritchey-Chr\'etien reflective telescope, Carl Zeiss JENA, Jena, German Democratic Republic (GDR / East Germany).}}
\end{table}

The daily data rates reported in Table~\ref{tab:telescopes} represent raw observational throughput prior to deterministic partitioning and indexing within the central FAI Data Lake. Previously accumulated legacy raw data are being prepared for systematic ingestion into KazVO in planned phases.

Observational proposals and target specifications are submitted via the KazVO web portal (\url{https://vo.fai.kz}, accessed on 3~November~2026.
), generating standardized job execution files. For automated telescope systems at ATO, these job files are queued directly to local control computers, whereas for other instruments, observations are carried out manually by station operators. {The acquired FITS frames are stored on local observatory servers.
Automated replication to the central FAI computing facility has been validated on pilot datasets, with routine deployment at the observing sites still in preparation. The central computing infrastructure is described in the following subsection.}

\subsubsection{Data Storage and Computing Facility} \label{subsubsec:servers}

To store, process, and host heterogeneous observational datasets and theoretical simulations, FAI maintains a dedicated computing and storage infrastructure. The hardware environment operates under Ubuntu Server Linux and is organized around a multi-node cluster connected to high-availability storage systems. Job scheduling and parallel computational workflows are managed via SLURM and MPI frameworks, supporting primary workloads such as $N$-body cosmological simulations and automated observational data processing pipelines. High-speed inter-node communication and external data access are supported by redundant network interfaces with load-balancing failover. Detailed technical specifications of the hardware cluster are summarized in Table~\ref{tab:cluster_specs}.

\begin{table}[H]
\caption{Hardware specifications and capacity of the FAI computing and storage cluster.}
\label{tab:cluster_specs}
\begin{adjustwidth}{-\extralength}{0cm}
\small
\begin{tabular*}{\fulllength}{@{\extracolsep{\fill}}ll@{}}
\toprule
\textbf{Parameter/Component} & \textbf{Specification/Value} \\
\midrule
Computing Nodes & 15 nodes (multi-core CPUs and dedicated GPUs) \\
Total CPU Capacity & 1268 cores/2536 threads \\
Total GPU CUDA Cores & 544,768 cores \\
Theoretical Peak Performance (FP32) & 127.2 TFLOPS (CPU)/2496.6 TFLOPS (GPU) \\
Storage Capacity & $>$$570$~TB (mirrored configuration)/$>$$100$~TB whole observational archive \\
Storage Infrastructure & 2$\times$ High-Availability NAS systems, 2$\times$ Management servers \\
Internal Network Interconnect & 10~Gb/s Ethernet, 400~Gb/s InfiniBand switch \\
External Connectivity & Dual redundant links (100~Mb/s fiber + 25~Mb/s radio-bridge, 125~Mb/s aggregated) \\
Operating Environment & Ubuntu Server Linux, SLURM workload manager, OpenMPI \\
\bottomrule
\end{tabular*}
\end{adjustwidth}
\end{table}

To prevent cascading failures and ensure independent maintenance of KazVO components, the software ecosystem is isolated within a Proxmox VE virtualization environment using four dedicated Linux Containers (LXC), Figure~\ref{fig:kazvo_arch}:

\begin{figure}[H]
\begin{adjustwidth}{-\extralength}{0cm}
\centering
\includegraphics[width=.9\linewidth]{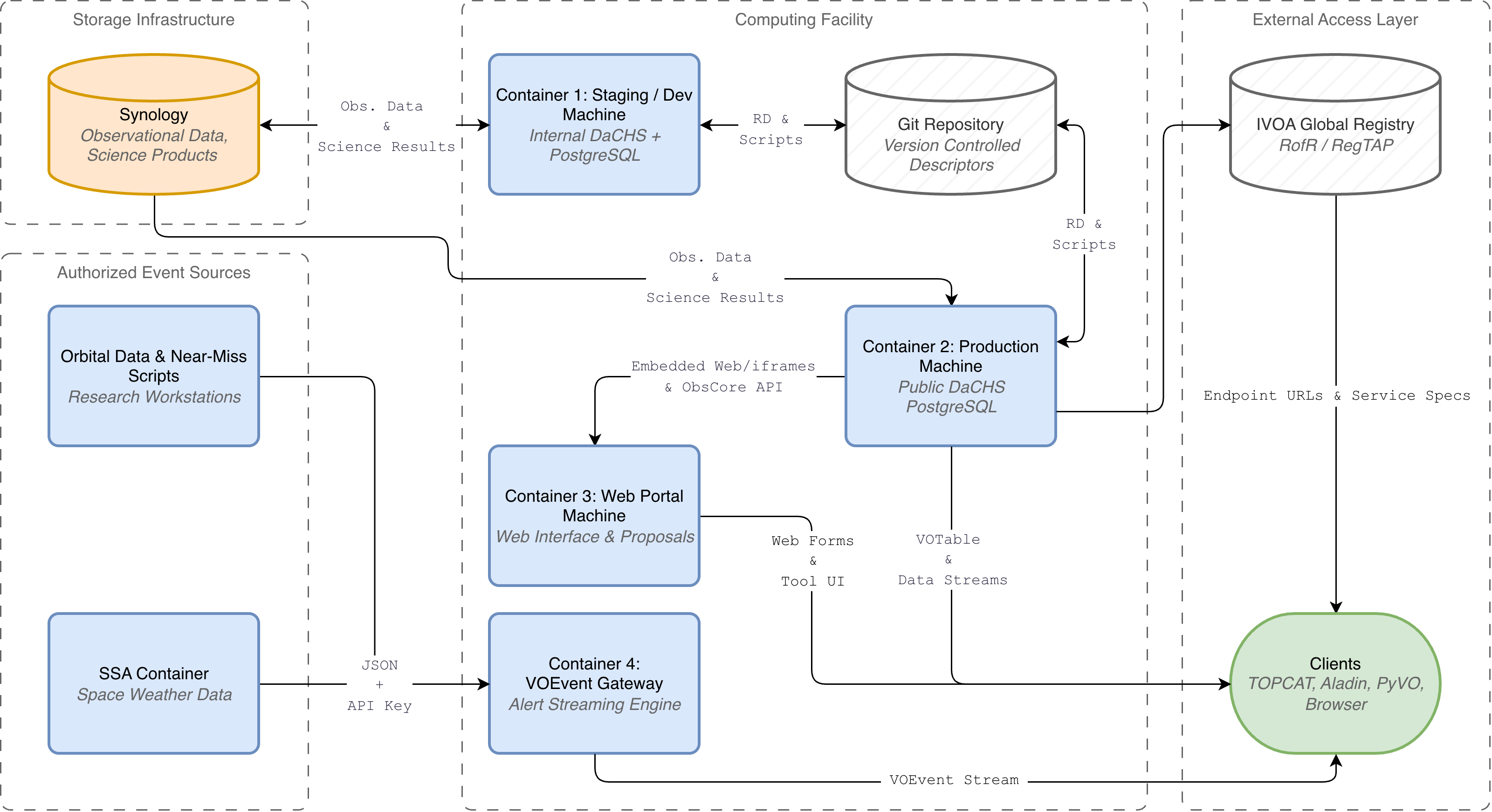}
\end{adjustwidth}
\caption{Containerized server-side architecture of the KazVO platform within the FAI computing facility.}
\label{fig:kazvo_arch}
\end{figure}

\begin{itemize}
\item DaCHS Staging (Container 1): an isolated internal environment running GAVO DaCHS~(v2.13) \citep{demleitner2014dachs} and PostgreSQL~(v15) \citep{postgresql15}, dedicated to resource descriptor debugging, script testing, and data model validation without risking operational services;

\item DaCHS Production (Container 2): the primary public KazVO node with an identical software stack, whose {publishing registry is registered in the IVOA Global Registry, i.e., Registry of Registries (RofR)} \citep{plante2007rofr}. It processes external {Machine-to-Machine} (M2M) queries via IVOA protocols and supplies embedded search interfaces to the web portal. Configuration changes are synchronized with the Staging environment via a Git (v2.47.3) repository \citep{git_kazvo_inputs}; 

\item Web Portal (Container 3): the central user-facing web interface for KazVO, providing interactive access to web tools, file downloads, proposal submissions, and incorporating embedded search views and forms from DaCHS Production;

\item VO Event Gateway (Container 4): an isolated microservice gateway running Ubuntu~22.04 LTS (FastAPI and Comet broker). The gateway accepts incoming real-time messages in JavaScript Object Notation (JSON) format with an API key verification from authorized internal sources, converts them into the VOEvent~2.0 (XML) standard, and streams the alert feed to external network subscribers.
\end{itemize}

Synology High Availability (HA) NAS, configured as a resilient two-node active--passive cluster, serves as the storage backbone housing the institutional Data Lake, the detailed file system hierarchy and partitioned directory structure are described in \linebreak  Section~\ref{subsubsec:storage}. The storage system consolidates calibrated FITS archives and derived science products, providing direct file system access exclusively to the DaCHS publishing backends (Containers~1~and~2). Adjacent system components, the Web Portal (Container~3) and the VOEvent Gateway (Container~4), operate in isolation, interacting with the platform directly via network API interfaces.

\subsection{Data Standardization, FAIR-ification, and Pipeline Architecture} \label{subsec:fair_pipeline}

{To support the FAIR principles across heterogeneous observational assets,} KazVO implements a multi-stage architecture encompassing archival digitization, structured storage partitioning, automated processing pipelines, and standardized IVOA publishing workflows.

\subsubsection{Digitization and Metadata Normalization}

Digitization of the FAI photographic archive aimed to save physical heritage assets for modern scientific use. On a technical level, emulsion optical density profiles from glass plates and film strips were converted into relative intensities, {and} handwritten observational logbooks were converted into machine-readable metadata. Scanning was conducted in transparency mode at an optimal spatial resolution of 1200~dpi (16-bit grayscale), which provides sufficient resolution of the emulsion grain structure at an uncompressed Tagged Image File Format (TIFF) file size of $\approx$$36\text{~MB}$. Detailed protocols for density calibration, characteristic curve construction, and geometric distortion correction for spectrograms are thoroughly documented in our previous works \citep{shomshekova2022na, shomshekova2022comets, shomshekova2023, izmailova2026, umirbayeva2025}.

To link physical negatives with digital headers, historical logbook records were structured into a local database. The TIFF scans were {converted into 16-bit FITS format} using Maxim~DL~Pro and Image Reduction and Analysis Facility (IRAF). Dedicated Python (v3.11) scripts generate FITS headers from the historical logbook metadata, converting recorded local time (LT) and local sidereal time (LST) to Coordinated Universal Time (UTC). The adopted convention uses the exposure midpoint for historical astroplates and the exposure start time for modern CCD frames. These conventions must be distinguished when interpreting temporal metadata. Metadata normalization and mandatory FITS header structure for digitized astroplates were implemented in strict accordance with IVOA \mbox{recommendations \citep{demleitner2022plates}}.

\subsubsection{Data Lake Partitioning and Ingestion Architecture} \label{subsubsec:storage}

A dedicated volume on the central Synology~HA~NAS (Section~\ref{subsubsec:servers}) is specifically allocated for the institutional Data Lake, consolidating both modern observational data streams from active telescopes and digitized heritage assets. The replication pipeline is architected to enable automated transfer of FITS datasets from remote observing facilities to the central node during scheduled off-peak hours, thereby avoiding network congestion and bandwidth contention during active nocturnal observations. Under the proposed operational model, raw frames generated by active instruments (Table~\ref{tab:telescopes}) are buffered on local observatory servers, undergo automated integrity verification, and are prepared for transmission via Starlink satellite terminals and terrestrial links; the inter-server synchronization workflow has been validated on pilot benchmark datasets and is currently undergoing preparation for field deployment.

To eliminate archive fragmentation and ensure high scalability, the physical structure of the institutional Data Lake employs a deterministic partitioning scheme (Partitioned Data Lake). The storage space is split into two primary domains: \texttt{/observations/} (observational datasets) and \texttt{/simulations/} (numerical models and synthetic spectra). The directory topology isolates assets by observatory, instrument, and observing mode, explicitly separating calibration frames from science targets for both modern digital detectors and historical digitized plates (Listing~\ref{lst:datalake_structure}).

\begin{listing}[H]
\caption{Canonical directory topology of the partitioned FAI Observational Data Lake.}
\label{lst:datalake_structure}
\rule{\columnwidth}{1pt}
\vspace{2pt}
\begin{tabular}{@{}l@{\hspace{1em}}l@{}}
\texttt{/fai/} & \\
\texttt{|-- observations/} & \\
\texttt{|\ \ `-- <observatory\_id>/} & \textcolor{gray}{\texttt{\# e.g., assy, tshao, kamenskoe}} \\
\texttt{|\ \ \ \ \ `-- <instrument\_id>/} & \textcolor{gray}{\texttt{\# e.g., azt20, zeiss1000\_east, schmidt}} \\
\texttt{|\ \ \ \ \ \ \ `-- <observ\_mode>/} & \textcolor{gray}{\texttt{\# e.g., ccd, spec\_slit, spec\_echelle}} \\
\texttt{|\ \ \ \ \ \ \ \ \ \ |-- calibration/} & \\
\texttt{|\ \ \ \ \ \ \ \ \ \ |\ \ `-- <YYYY>/} & \\
\texttt{|\ \ \ \ \ \ \ \ \ \ |\ \ \ \ `-- <YYYY-MM-DD>/} & \textcolor{gray}{\texttt{\# bias/, dark/, flat/, lamp/, masters/}} \\
\texttt{|\ \ \ \ \ \ \ \ \ \ `-- targets/} & \\
\texttt{|\ \ \ \ \ \ \ \ \ \ \ \ \ `-- <YYYY>/} &\\
\texttt{|\ \ \ \ \ \ \ \ \ \ \ \ \ \ \ `-- <YYYY-MM-DD>/} & \textcolor{gray}{\texttt{\# raw/, pre-reduced/, reduced/, meta/}} \\
\texttt{`-- simulations/} & 
\end{tabular}
\vspace{2pt}
\rule{\columnwidth}{1pt}
\end{listing}

In contrast to bulk FITS images and simulation snapshots residing on the Synology~HA~NAS, standalone tabular datasets, published electronic catalogs, and curated VO service tables (described in detail in Section~\ref{subsec:res_holdings}) are maintained natively on the Production DaCHS node in Container~2 described in Section~\ref{subsubsec:servers}. These assets are stored within the internal DaCHS resource directory structure (\texttt{/var/gavo/inputs/}) and ingested directly into the local relational PostgreSQL database, optimizing query execution speed and TAP protocol responsiveness for non-image services.

\begin{figure}[H]
\includegraphics[width=0.95\textwidth]{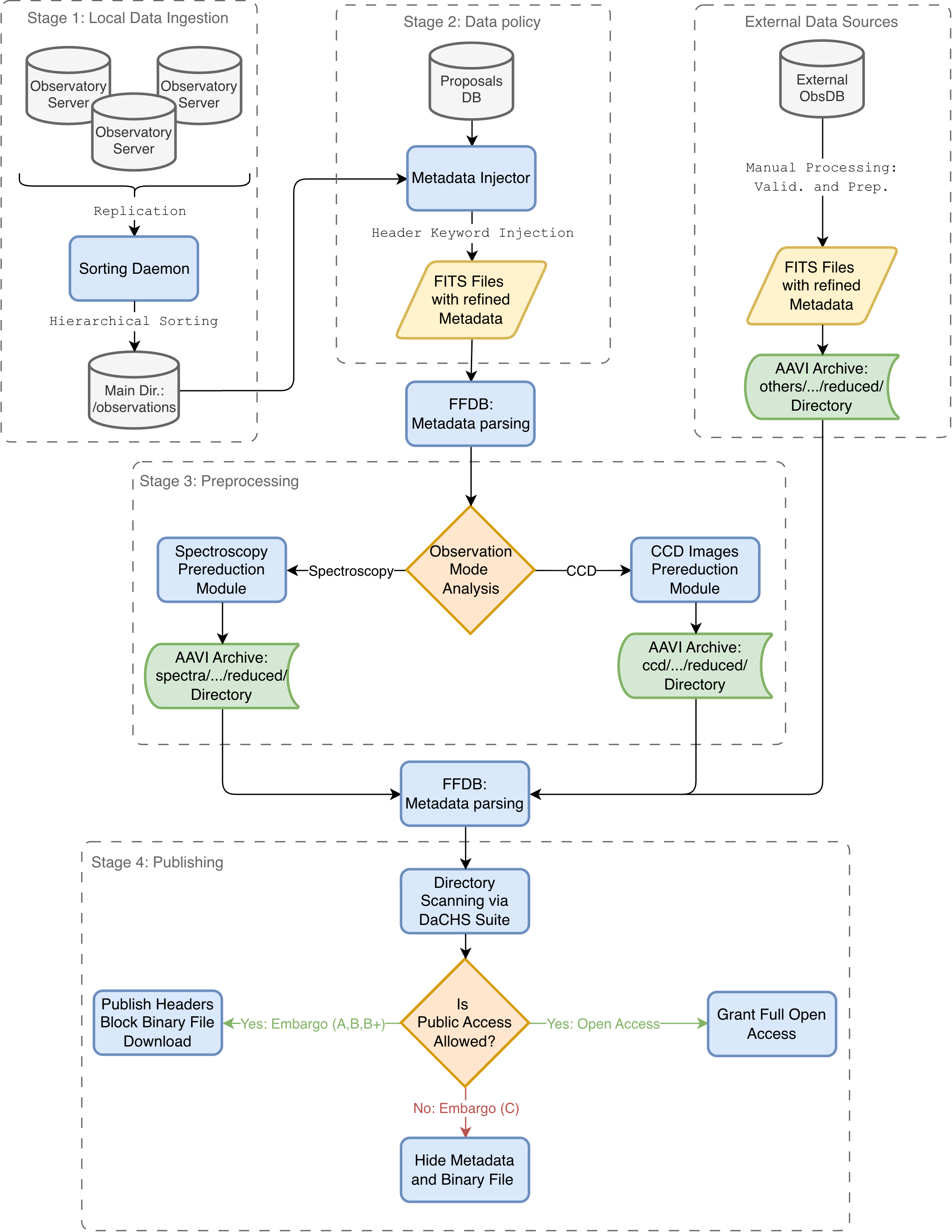}
\caption{Architecture of the KazVO {modular} ingestion, pre-processing, and publishing pipeline. The diagram illustrates the sequential progression of FITS datasets through server replication (Stage~1), embargo metadata injection (Stage~2), mode-specific pre-processing (Stage~3), and integration with the DaCHS backend to realize the Metadata Release (``Scientific Showcase'') model (see Section~\ref{subsec:reg_datapolicy}).}
\label{fig:ingestion_pipeline}
\end{figure}

Sorting of incoming FITS datasets across the directory hierarchy is implemented as an ingestion software module operating in a scheduled, batch-ingestion mode. {The modular workflow encompassing data ingestion, initial reduction, and metadata publication (Figure~\ref{fig:ingestion_pipeline}) currently operates in batch mode as observational datasets accumulate, while automated site-to-center replication (Stage~1) has been validated on pilot datasets.}

During pipeline processing, files undergo validation against the KazVO Standard FITS Header Keywords specification \citep{kazvo_fits_keywords}. The Metadata Injector module enriches headers with data policy keywords, e.g., \texttt{OBS\_CLSS}, \texttt{RELEASE}, \texttt{PROPOSID}, the detailed data access policy and embargo framework are described in Section~\ref{subsec:reg_datapolicy}, linking each dataset to the corresponding proposal database entry. Following the applicable reduction steps, processed FITS products are stored in the \texttt{reduced/} directory by observation date. An incremental metadata import into GAVO DaCHS then makes the designated products discoverable through IVOA endpoints. Header enrichment and metadata import do not, by themselves, change the ObsCore calibration level.

\subsubsection{Frames Calibration and Dual Metadata Indexing} \label{subsubsec:reduction}

\begin{figure}[H]
\includegraphics[width=0.55\linewidth]{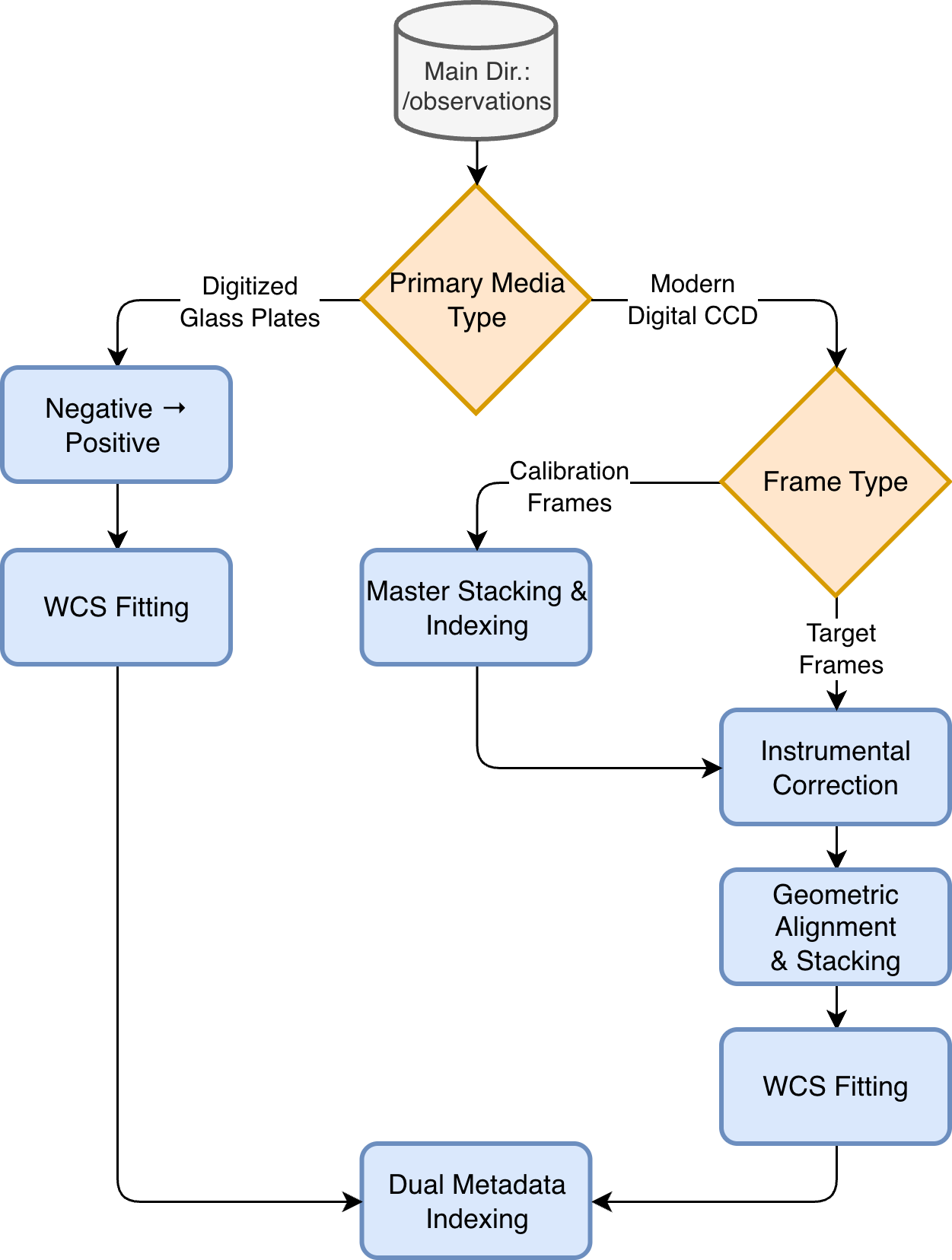}
\caption{Operational flowchart of pre-reduction pipeline for observational photometric data streams.}
\label{fig:prered_pipeline}
\end{figure}

Following ingestion and header enrichment (Section~\ref{subsubsec:storage}), raw FITS frames residing in the Data Lake enter the automated pre-reduction pipeline (Figure~\ref{fig:prered_pipeline}). {The pipeline distinguishes instrumental FITS inputs (Level~1), calibrated individual science products (Level~2), and enhanced products formed by alignment and co-addition of science exposures (Level~3). These categories describe processing stages and do not constitute an independent assessment of scientific accuracy.}

For observational photometric assets, the processing workflow explicitly differentiates between primary physical media:
\begin{itemize}
    \item \textbf{Modern CCD Streams:} Digital frames undergo automated {dark} subtraction and {flat}-field normalization using pre-indexed calibration databases, followed by relative geometric alignment and robust sigma-clipped co-addition into Master Stacks. Targeted WCS astrometric calibration is subsequently fitted directly on the synthesized Master Stack using a local \textit{Astrometry.net} solver \citep{lang2010, astrometry_index_url} with {distortion corrections represented using the Simple Imaging Polynomial (SIP) convention}. Primary reduction and co-addition for these streams are implemented via a custom resource-bounded Python pipeline built upon Astropy \citep{astropy:2013,astropy:2018,astropy:2022} and \texttt{ccdproc} \citep{matt_craig_2017_1069648}, with scripts openly hosted in our institutional GitHub repository \citep{fai_sort_repo}.
    
    \item \textbf{Digitized Historical Archive:}
    Photographic plates undergo grayscale negative-to-positive transformation, followed by automated source detection \citep{1996A&AS..117..393B, Barbary2016} and WCS solution. For the historical plates included in the residual analysis in Section~\ref{subsec:res_performance}, the FITS processing records identify Tycho-2-based astrometric indices. The subsequent comparison with Gaia DR3 and its methodological limitations are described in that section.
\end{itemize}

The architecture and decision logic of this photometric reduction pipeline are schematically illustrated in Figure~\ref{fig:prered_pipeline}.

Upon completion of primary calibration and WCS fitting, datasets enter parallel metadata indexing streams. Internal resource accounting and comprehensive archive tracking are managed by the {FITS Header Database Service (FFDB, \citep{ffdb_repo})}, which parses headers across all Data Lake levels, including raw frames, calibration sets, and embargoed files, to maintain an administrative inventory for fast internal queries. In parallel, the GAVO DaCHS publishing backend serves as the external VO gateway: {it parses headers from the observational products selected for publication and declared in Resource Descriptors} and populates the relational PostgreSQL database. This separation guarantees that DaCHS exposes only authorized public metadata via standardized IVOA protocols, while FFDB preserves complete administrative control over institutional datasets.

While the automated pipeline shown in Figure~\ref{fig:prered_pipeline} currently covers {targeted} photometric observations, spectroscopic data streams are handled separately. Digitized historical slit spectra require interactive reduction using dedicated software tools \citep{izmailova2026} and IRAF due to complex dispersion profiles, whereas fully automated pipelines for modern spectroscopic data are being developed in stages \citep{gluchshenko2026}. Data processing for specialized dynamic surveys is carried out by project investigators under tailored agreements.
\subsubsection{IVOA Interoperability Layer and Standards Compliance} \label{subsubsec:ivoa_layer}

\begin{figure}[H]

\includegraphics[width=0.95\textwidth]{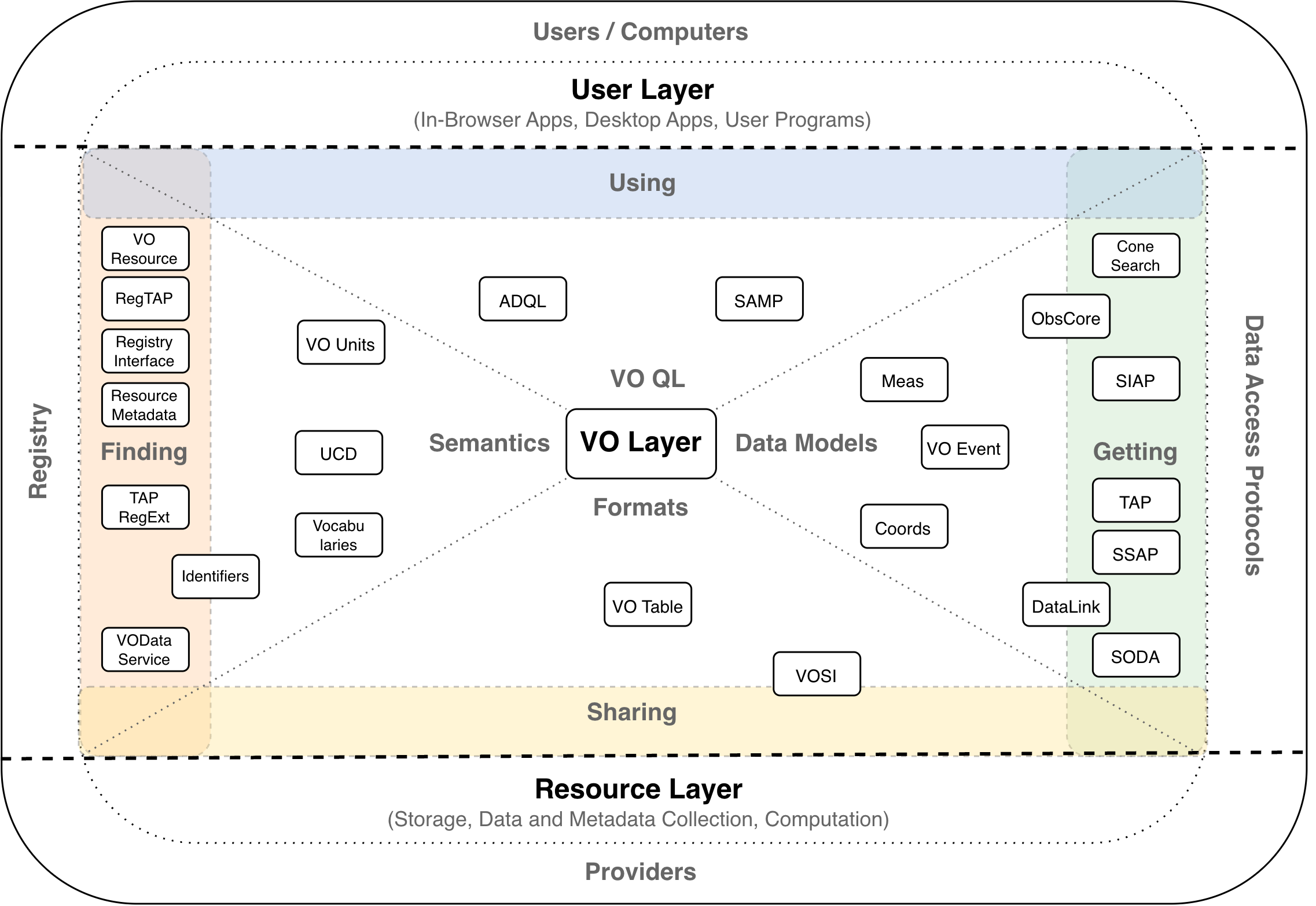}
\caption{The architecture of KazVO mapped onto the IVOA reference framework.}
\label{fig:kazvo_ivoa_arch}
\end{figure}

The integration of KazVO into the IVOA ecosystem strictly complies with the IVOA architecture (Figure~\ref{fig:kazvo_ivoa_arch}), covering the entire data lifecycle: from the hardware-resource layer (\textit{Resource Layer}) to end-user applications (\textit{User Layer}).

The interoperability layer bridging hardware resources and end-users is realized via core IVOA components (\textit{VO Layer}), ensuring metadata standardization and semantic coherence. Metadata extracted from processed FITS headers are mapped to the mandatory relational schema of the \textbf{ObsCore (v1.1)} data model \citep{ivoa.obscore2017} during the construction of GAVO DaCHS resource descriptors (\texttt{q.rd}). Through this mapping, metadata fields are structured according to the \textbf{Coordinates} \citep{rots2022coords} and \textbf{Measurements} \citep{rots2022meas} data models, and are annotated with Unified Content Descriptors (\textbf{UCD} \citep{genova2019ucd}), standardized physical units (\textbf{VOUnits} \citep{derriere2014vounits}), and controlled terms from \textbf{IVOA Vocabularies} \citep{demleitner2023vocabularies}. Metadata query responses are structured in the XML-based \textbf{VOTable} format \citep{ochsenbein2019votable}, while programmatic database querying is driven by the Astronomical Data Query Language (\textbf{ADQL} \citep{osuna2008adql}).

The functional access, discovery, and utilization layer is structured around key IVOA capabilities \citep{dowler2021arch}:
\begin{itemize}
\item Getting (Data Access Protocols): providing M2M access to relational tables, photometric images, and spectra via TAP (v1.1), SIAP (v2.0), SSAP (v1.1), and Cone \mbox{Search \citep{plante2008scs}} protocols. Delivery of associated files and on-the-fly spatial/spectral cutouts is provided via DataLink (v1.1) \citep{bonnarel2023datalink} and Server-side Operations for Data Access (SODA) (v1.0) \citep{bonnarel2017soda} services managed by the DaCHS backend, while real-time alert streaming is handled via VOEvent (v2.0) \citep{swinbank2017vtp} through a dedicated gateway.

{\textbf{Finding (Registry Services):} KazVO describes its resources using VOResource metadata and IVOA identifiers (IVOIDs) \citep{demleitner2016identifiers}. The KazVO publishing registry is registered in the IVOA Registry of Registries (RofR). Its resource records are harvested by searchable registries, allowing clients to discover the associated services through registry queries, including RegTAP \citep{demleitner2019regtap} and VOResource metadata descriptions \citep{plante2018voresource} (including the Table Access Protocol Registry Extension (TAPRegExt) \citep{demleitner2012tapregext}).}
\item Sharing (Infrastructure Services): exposing service status metadata, availability parameters, and table schemas via native IVOA Support Interfaces (VOSI) \citep{graham2017vosi} endpoints provided by DaCHS.
\item Using (User Applications \& Client Libraries): direct integration of published services with desktop applications (TOPCAT \citep{taylor2017topcat}, Aladin \citep{bonnarel2000aladin}), programmatic access libraries (PyVO \citep{graham2014pyvo}, Astropy), and the native KazVO web portal, with inter-application messaging and seamless data exchange enabled via the Simple Application Messaging Protocol (SAMP) \citep{boch2010samp}.
\end{itemize}

{Together, these IVOA components support standardized discovery, access, and use of the published KazVO resources by external applications and clients. Their contribution to the FAIR principles is summarized in Section~\ref{subsubsec:fair_mapping}.}

\subsubsection{Global Registry Discovery and Data Policy} \label{subsec:reg_datapolicy}

Compliance with the FAIR principles of Findability and Accessibility is supported through the registration of {the KazVO publishing registry} within the global IVOA registry ecosystem via the RofR. This makes FAI observational data findable and accessible for M2M queries executed by external client applications, such as TOPCAT, Aladin, and the PyVO library. The institutional data governance framework is grounded in the ``KazVO Observation Data Use Policy and License Agreement''~\citep{kazvo_policy}, formulated in alignment with the best practices from astronomical data centers, including ESO~\citep{eso_policy}, HST~\citep{hst_policy}, Chandra~\citep{chandra_policy}, National Radio Astronomy Observatory~(NRAO)~\citep{nrao_policy}, and Keck Observatory \citep{keck}.

\begin{figure}[H]
\includegraphics[width=0.95\textwidth]{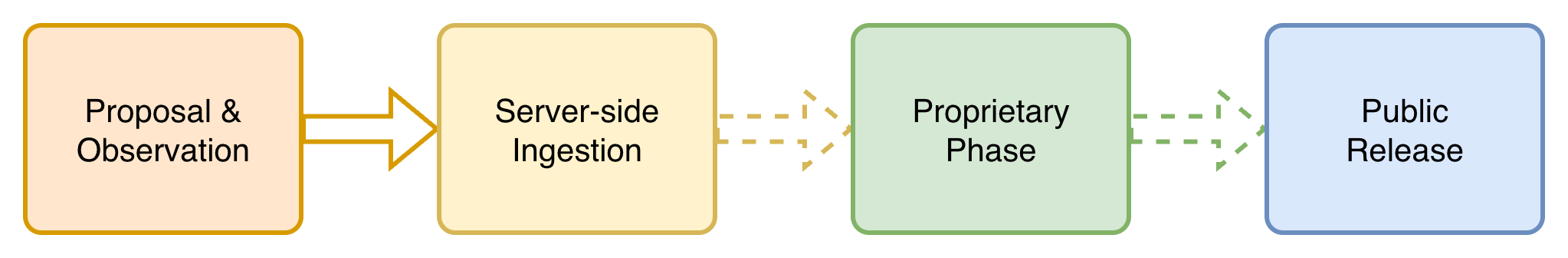}
\caption{Schematic representation of the KazVO data lifecycle and access control framework. Solid arrows indicate the principal processing steps in the depicted workflow; dashed arrows indicate conditional transitions governed by the applicable access policy. The diagram describes the workflow and does not imply that every step has been applied to every archived FITS file.}

\label{fig:data_lifecycle_concept}
\end{figure}

The {overall} data lifecycle, spanning from observational proposal submission via the web portal to automated embargo expiration, is illustrated in Figure~\ref{fig:data_lifecycle_concept}.

The policy strictly regulates the legal status of institutional assets. 
Historical heritage from the ``Glass Library'' plate archives (Maksutov ASI-2 and Schmidt cameras, 1950--1997) is dedicated to the Public Domain under the Creative Commons Public Domain Dedication (\textbf{CC0 1.0}), reflecting their status as digitized physical heritage and scientific raw facts where modern copyright restrictions do not apply.
For modern digital streams and processed science-ready products generated by active telescopes {(such as CCD photometry and extracted spectra)}, a flexible 4-tier access classification scheme (Classes A, B, B+, C) with automated lifecycle management is implemented (Table~\ref{tab:kazvo_data_classes}). The standardized Creative Commons Attribution license (\textbf{CC BY 4.0}) serves as the default license for open-access {modern datasets}, ensuring mandatory citation of the observatory and principal investigators in all derivative research.

\begin{table}[H]
\caption{Classification of observational datasets and access modes in KazVO.}
\label{tab:kazvo_data_classes}
\begin{adjustwidth}{-\extralength}{0cm}
\small
\begin{tabular*}{\fulllength}{@{\extracolsep{\fill}}llll@{}}
\toprule
\textbf{Class} & \textbf{Data Category} & \textbf{Embargo Period} & \textbf{Post-Embargo KazVO Status} \\
\midrule
\textbf{A} & Standard scientific observations & 12 months (default) & Public \\
\textbf{B} & Monitoring programs & Up to 36 months & Public \\
\textbf{B+} & Long-term surveys and collaborations & Up to 60 months & Public \\
\textbf{C} & Strategic/Commercial datasets & Perpetual/NDA & Restricted/Unlisted \\
\bottomrule
\end{tabular*}
\end{adjustwidth}
\end{table}

A core element of our policy is the ``Scientific Showcase'' (Metadata Release) model, whereby observational metadata (FITS headers) become queryable immediately upon ingestion, even during the active proprietary embargo period. This informs the scientific community about ongoing research programs and prevents redundant duplicate observations. To technically enforce this policy, the server ingestion pipeline (Section~\ref{subsubsec:storage}) automatically injects a standardized set of governance keywords into the primary FITS headers (Table~\ref{tab:fits_governance_keywords}).

\begin{table}[H]
\caption{Specification of FITS governance, access control, and citation keywords in KazVO.}
\label{tab:fits_governance_keywords}
\begin{adjustwidth}{-\extralength}{0cm}
\footnotesize
\begin{tabular*}{\fulllength}{@{\extracolsep{\fill}}lll@{}}
\toprule
\textbf{FITS Keyword} & \textbf{Description/Purpose} & \textbf{Data Source} \\
\midrule
\texttt{PI\_NAME}  & Principal Investigator (PI) full name & Applicant user account \\
\texttt{CO\_I}     & Responsible Co-Investigator (Co-I) & Application field \\
\texttt{PROPOSID} & Unique Observation Identifier (UOID) & Submission ID \\
\texttt{OBS\_CLSS} & Access category (Classes A, B, B+, C) & Proposal classification \\
\texttt{RELEASE}  & Public release date (ISO 8601 \citep{iso8601}) & Observation date + Embargo period \\
\texttt{CONTACT}  & Public e-mail of PI for collaboration & PI contact details \\
\texttt{LICENSE}  & Applied data license (\texttt{CC0-1.0} for historical plates or \texttt{CC-BY-4.0} for modern products) & Collection data-use policy \\
\bottomrule
\end{tabular*}
\end{adjustwidth}
\end{table}

{For the CCD archive, discovery metadata remain queryable during embargo, while access to the original FITS data is controlled by the DaCHS product infrastructure and the dedicated \texttt{zeroed} delivery service. The delivery handler reads the product's embargo date from \texttt{dc.products} and selects either the masked representation or the original file according to the current server date. This request-time delivery check is distinct from SQL date filters used to select records in the browsing interface. The release logic and its controlled validation are described in Section~\ref{subsec:embargo_enforcement}.}

\subsubsection{{Mapping to the FAIR Principles and Resource Identification}}
\label{subsubsec:fair_mapping}

The KazVO architecture supports the FAIR principles through
a combination of resource discovery, standardized data access,
shared metadata conventions, and documented conditions for reuse.
These mechanisms connect the infrastructure described above
to the practical requirements of finding, accessing, interpreting,
and reusing heterogeneous astronomical data.
Table~\ref{tab:fair_mapping} summarizes this correspondence
\citep{wilkinson2016fair}, indicating the scope and limitations
of the current implementation.

\begin{table}[H]
\caption{{Mapping of the FAIR principles to KazVO mechanisms and the scope of the current implementation.}}
\label{tab:fair_mapping}
\small

\begin{adjustwidth}{-\extralength}{0cm}
\begin{tabular*}{\fulllength}{@{}p{0.15\textwidth}p{0.52\textwidth}p{0.56\textwidth}@{}}
\toprule
\textbf{Principle} &
\textbf{KazVO Mechanisms} &
\textbf{Scope and Limitations} \\
\midrule

Findable &
Registered resources are identified by IVOIDs and described through
VO Registry metadata. Published observational holdings can be
searched through their metadata using TAP/ObsCore and the relevant
image or spectral discovery services. &
Registry identifiers apply to registered resources; their presence
does not imply a separate persistent identifier for every underlying
file or a version-specific identifier for each collection release. \\
\midrule

Accessible &
Published data and metadata are exposed through standard IVOA
interfaces. For holdings covered by the Scientific Showcase model,
discovery metadata remain accessible during embargo, while access
to the original data arrays is restricted. &
Access depends on the applicable data policy.
Long-term preservation of metadata independently of
data availability is not evaluated in this study. \\
\midrule

Interoperable &
FITS and VOTable provide standardized data and metadata formats.
ObsCore, UCDs, VOUnits, and IVOA vocabularies provide shared
structures and semantics for the relevant published resources. &
The applicable data model and protocol depend on the resource type.
ObsCore describes observational products and is not a universal
schema for all catalogs, simulation grids, or interactive tools. \\
\midrule

Reusable &
Observational metadata, descriptions of instruments and processing
methods, and the institutional data-use policy provide context for
interpretation, attribution, and reuse. &
Reuse conditions depend on the collection and its applicable
license. Processing and provenance information should be assessed
for the particular product; its completeness is not established
uniformly across all holdings in this study. \\

\bottomrule
\end{tabular*}
\end{adjustwidth}
\end{table}

KazVO currently uses IVOA identifiers (IVOIDs) to identify registered resources within the Virtual Observatory infrastructure. Resource-level identification should be distinguished from the identification of individual files or immutable, versioned dataset releases. KazVO does not currently assign DOIs to its collections or services, and no DOI assignment workflow is planned at this stage. DOI-based citation of versioned releases may be considered as a future extension.



\section{Results} \label{sec:results}

{This section describes the current implementation and published holdings of KazVO. It reports exploratory astrometric residual measurements, examples of data access, and controlled tests of the embargo-delivery mechanism, with the scope and limitations of each assessment stated below.}

\subsection{Data Lake Topology and Storage Structure} \label{subsec:datalake_quantification}

The KazVO Data Lake, deployed on a high-availability Synology HA NAS storage array ($>$$570$~TB capacity), consolidates heterogeneous scientific datasets from primary FAI observatories and partner facilities. As detailed in Section~\ref{subsubsec:storage} the Data Lake architecture is logically partitioned by observatory site, observational modality, and historical epoch (Table~\ref{tab:datalake_summary}).

{Table~\ref{tab:datalake_summary} summarizes the broader Data Lake inventory, including raw observations, calibration files, logs, intermediate products, and processed data. The listed volumes sum to approximately 9.70~TB. These inventory values should be distinguished from the production-service audit of 28 October 2026 in Table~\ref{tab:kazvo_holdings}, which covers a narrower selection of observational products, catalog and monitoring ingestion files, and model files. The two inventories differ in scope and are not a synchronized, file-by-file reconciliation. Their volumes must not be added together.}

\begin{table}[H]
\caption{{Structure, temporal coverage, and reported file volumes of the integrated observational holdings in the KazVO Data Lake. The inventory includes raw observations, calibration files, intermediate products, logs, and processed products. Volumes describe the listed holdings, not the allocated storage of the NAS including replication, snapshots, or filesystem overhead.}}
\label{tab:datalake_summary}
\begin{adjustwidth}{-\extralength}{0cm}
\begin{tabularx}{\fulllength}{@{} 
  >{\raggedright\arraybackslash}X 
  >{\raggedright\arraybackslash}p{0.25\fulllength} 
  c 
  >{\raggedleft\arraybackslash}p{0.18\fulllength} 
  >{\raggedleft\arraybackslash}p{0.13\fulllength} 
@{}}
\toprule
\textbf{Instrument \textsuperscript{1}} & \textbf{Observation Mode} & \textbf{Time Span} & \textbf{Data} & \textbf{Volume} \\
{} & {} & \textbf{[yr]} & \textbf{[Files]} & \textbf{[GB]} \\
\midrule
\multicolumn{5}{l}\textit{1. Assy-Turgen Observatory} \\
\multirow{2}{*}{AZT-20} & Slit Spectroscopy & 2021--2026 & 1110  & 3.57  \\
 & CCD Photometry & 2020-- & 63,260  & 460.7 \\
WFOS-70 & Wide-Field CMOS Phot. & {2025}-- & 811 & 50.95 \\
WFOS-40 & Wide-Field CMOS Phot. & 2024-- & 133,094 & 3493.1 \\
RC-500 & Photometry (CMOS) & 2021--2022 & 2101 & 77.24 \\
\midrule
\multicolumn{5}{l}{\textit{2. Tien Shan Astronomical Observatory}} \\
Zeiss-1000 (East) & CCD Photometry & 2013-- & 463,972 & 4036.17  \\
Zeiss-1000 (West) & Slit Spectroscopy & 2014--2021 & 83,564 & 475.68 \\
\midrule
\multicolumn{5}{l}{\textit{3. Kamenskoe Plateau Observatory}} \\
ASI-2  & Digitized Glass Plates & 1950--1997 & 21,805 & 542.51 \\
Schmidt Camera & Digitized Glass Plates & 1964--1989 & 8304 & 253.55 \\
\multirow{2}{*}{AZT-8 Telescope} & Digitized Films & {1970--1997} & 36,812  &  278.6  \\
 & CCD Slit Spec. & 2003-- & 7809  & 23.6  \\
\midrule
\multicolumn{5}{l}{\textit{4. Collaborative Facilities \& Solar Observatories}} \\
Eshel-TCO & Echelle Spectroscopy & 2013--2024 & {837} & 0.137 \\
{Space Weather Stations} &
{Time-Series CSV Files} &
2022-- &
{4136} &
{0.3684} \\
\midrule
\textbf{Total Integrated Dataset} & \textbf{All Modalities Combined} & \textbf{1950--} & \begin{tabular}[t]{@{}r@{}} 827,615  \end{tabular} & \begin{tabular}[t]{@{}r@{}} 9696.12 \end{tabular} \\
\bottomrule
\end{tabularx}

\vspace{2mm}

\end{adjustwidth}
\noindent{\footnotesize{\textsuperscript{1}~
\textbf{AZT-20}: ``Astronomicheskiy Zerkal'nyy Teleskop'' model No.~20, LOMO, Leningrad, USSR; 
\textbf{WFOS-70}: 70-cm aperture Wide-Field Optical System, AstroTechPribor, Almaty, Kazakhstan;  
\textbf{WFOS-40}: 40-cm aperture Wide-Field Optical System, AstroTechPribor, Almaty, Kazakhstan;
\textbf{RC-500}: 500-mm aperture Ritchey-Chr\'etien reflective telescope, AstroSib, Novosibirsk, Russia;
\textbf{Zeiss-1000 (East)}: ``East'' 1-m aperture Ritchey-Chr\'etien reflective telescope, Carl Zeiss JENA, Jena, German Democratic Republic (GDR / East Germany); 
\textbf{Zeiss-1000 (West)}: ``West'' 1-m aperture Ritchey-Chr\'etien reflective telescope, Carl Zeiss JENA, Jena, German Democratic Republic (GDR / East Germany);
\textbf{ASI-2}: ``Astronomicheskiy Spektral'nyy Teleskop'' model No.~2, also known as Maksutov Telescope, Gosudarstvennyy Opticheskiy Institut (Vavilov State Optical Institute), Leningrad, USSR;
\textbf{Schmidt}: Schmidt Camera (Catadioptric Astrophotographic Telescope), Carl Zeiss JENA, Jena, German Democratic Republic (GDR / East Germany);
\textbf{AZT-8}: ``Astronomicheskiy Zerkal'nyy Teleskop'' model No.~8, LOMO, Leningrad, USSR;
\textbf{Eshel-TCO}: eShel Fiber-Fed Echelle Spectrograph (Shelyak Instruments, Le Versoud, France) coupled to the 0.81-m telescope (Sigma Research Corporation, Richland, WA, USA) at the Three College Observatory, North Carolina, USA;
\textbf{Space Weather Stations}: \textbf{Space Weather Stations}: facilities operated by the
Institute of Ionosphere, Almaty, Kazakhstan:
(i) Alma-Ata Geomagnetic Observatory (IAGA code AAA),
equipped with LEMI-008 and LEMI-018 fluxgate magnetometers,
a POS-1 proton magnetometer, and a LEMI-203 fluxgate
declinometer mounted on a 3T2KP theodolite;
(ii) Orbita Radio Observatory (2700~m above sea level),
equipped with a 12-m parabolic antenna for solar radio
observations at 2.8~GHz and a CALLISTO radio spectrograph
covering 45--870~MHz;
(iii) Alma-Ata Cosmic Ray Station (AATB; 3340~m above sea level),
equipped with an 18NM-64 neutron supermonitor for continuous
cosmic-ray monitoring with one-minute accumulation intervals.}}
\end{table}

\subsection{{Quality Control and Astrometric
Residual Assessment}}
\label{subsec:res_performance}

{An exploratory assessment of positional agreement with Gaia DR3 was performed for selected processed images. The displayed results use the first 30 frames with successful residual measurements in each analysis group. This selection was not designed to provide a statistically representative sample of the archive.}

\begin{figure}[H]
\begin{adjustwidth}{-\extralength}{0cm}
\centering
\subfloat[\centering]{%
\includegraphics[width=9.0cm]{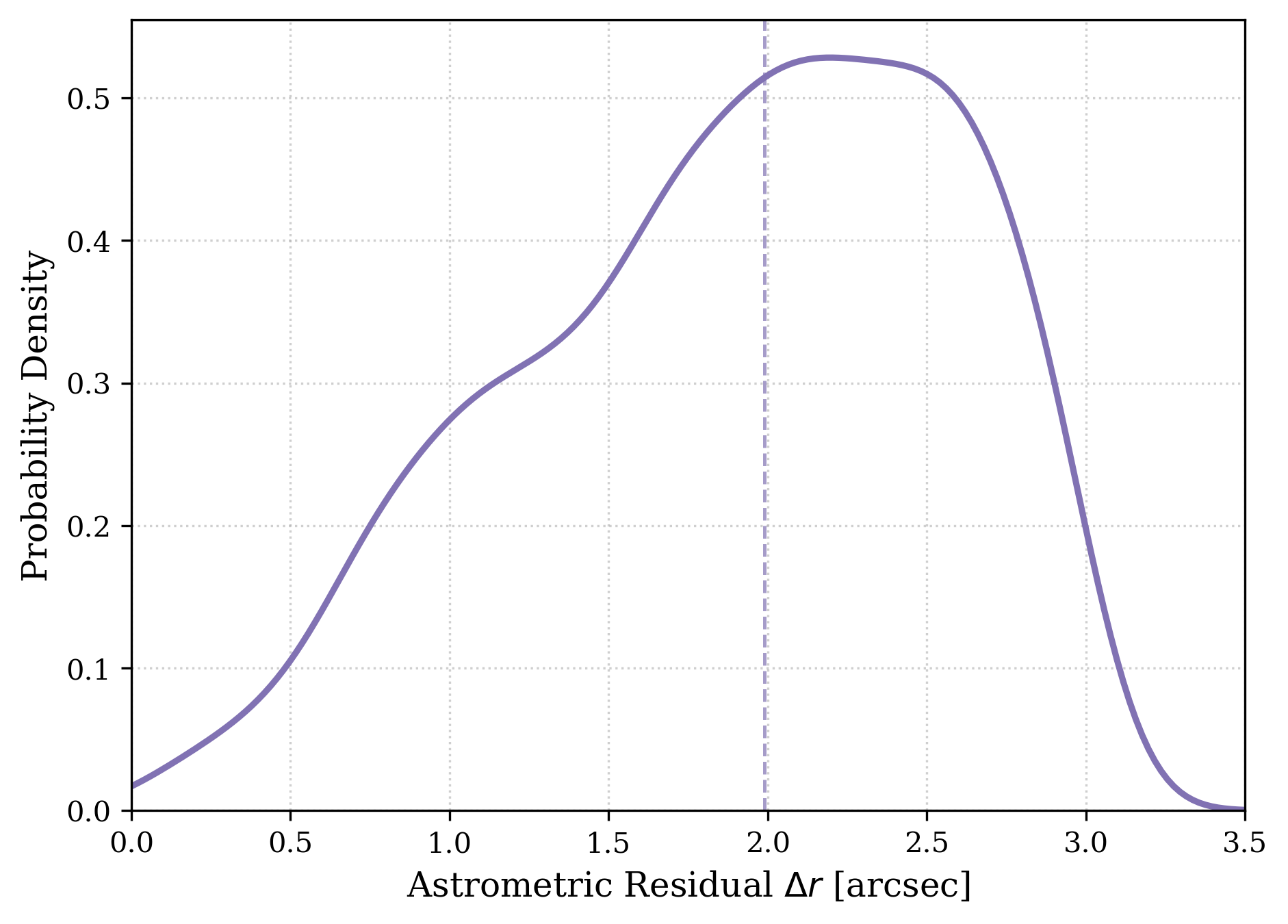}}
\hfill
\subfloat[\centering]{%
\includegraphics[trim={0 0 0 20}, clip, width=9.0cm]{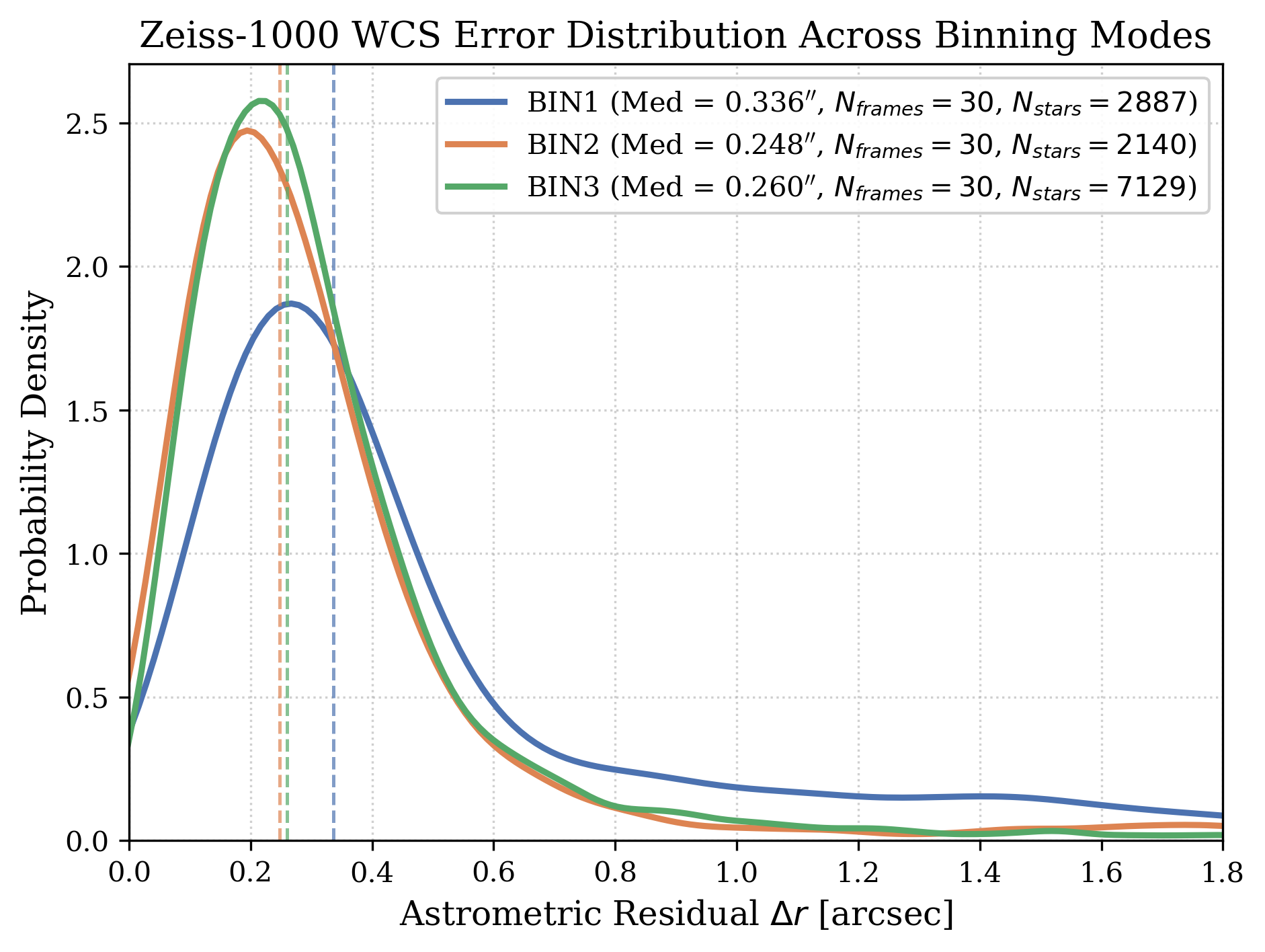}}
\end{adjustwidth}
\caption{{Distributions of radial on-sky positional residuals ($\Delta r$) relative to Gaia DR3 for accepted source matches: (\textbf{a}) digitized photographic plates from the 50-cm Maksutov meniscus telescope (ASI-2), with 30 frames, 1182 matches, and a median residual of $1.988''$; (\textbf{b}) CCD images from the 1-m Zeiss-1000 telescope, with 30 frames per binning mode (\texttt{BIN1}, \texttt{BIN2}, and \texttt{BIN3}). Vertical dashed lines indicate the corresponding median residual values for each group. Each group comprises the first 30 frames with successful residual measurements in the analysis. Gaia DR3 positions were not propagated to the observation epochs, and no explicit stellar hold-out subset was used.}}
\label{fig:wcs_validation}
\end{figure}

{For CCD images acquired with the 1-m Zeiss-1000 telescope, the median radial residuals of the accepted source matches are
$0.336''$ ($N_{\text{matches}} = 2887$) for \texttt{BIN1}, $0.248''$ ($N_{\text{matches}} = 2140$) for \texttt{BIN2}, and $0.260''$ ($N_{\text{matches}} = 7129$) for \texttt{BIN3}, using 30 frames per binning mode
\textls[-15]{(Figure~\ref{fig:wcs_validation}b). For digitized photographic plates from the 50-cm Maksutov meniscus telescope (ASI-2), the corresponding median radial residual is $1.988''$, based on 1182 accepted source matches across 30 frames (Figure~\ref{fig:wcs_validation}a). These values summarize the pooled source matches in each group, rather than the distribution of per-frame median residuals.}}

{The FITS processing records identify Tycho-2-based astrometric indices for the historical plates included in this analysis and Gaia-based indices for the CCD frames. The plate residuals therefore provide a comparison against a different reference catalog. However, no explicit stellar hold-out subset was reserved, and the CCD residuals should not be interpreted as an independent validation of astrometric accuracy.}

{Gaia DR3 positions were used without propagation to the observation epochs. Proper motions can therefore contribute materially to the residuals for historical plates. For illustration, a proper motion of 20~mas\,yr$^{-1}$ corresponds to a displacement of 1.2~arcsec over 60~years; this example is not a measured correction for the analyzed sample.}

These results characterize positional agreement for the analyzed frames and accepted source matches only. They do not establish archive-wide astrometric accuracy or validate the scientific suitability of all published products. Independent validation of the CCD solutions, propagation of reference positions to the observation epochs, and a broader assessment of sample representativeness remain necessary for such conclusions.

\subsection{Heterogeneous Data Integration} \label{subsec:res_verification}
\begin{figure}[H]
\includegraphics[width=1\textwidth]{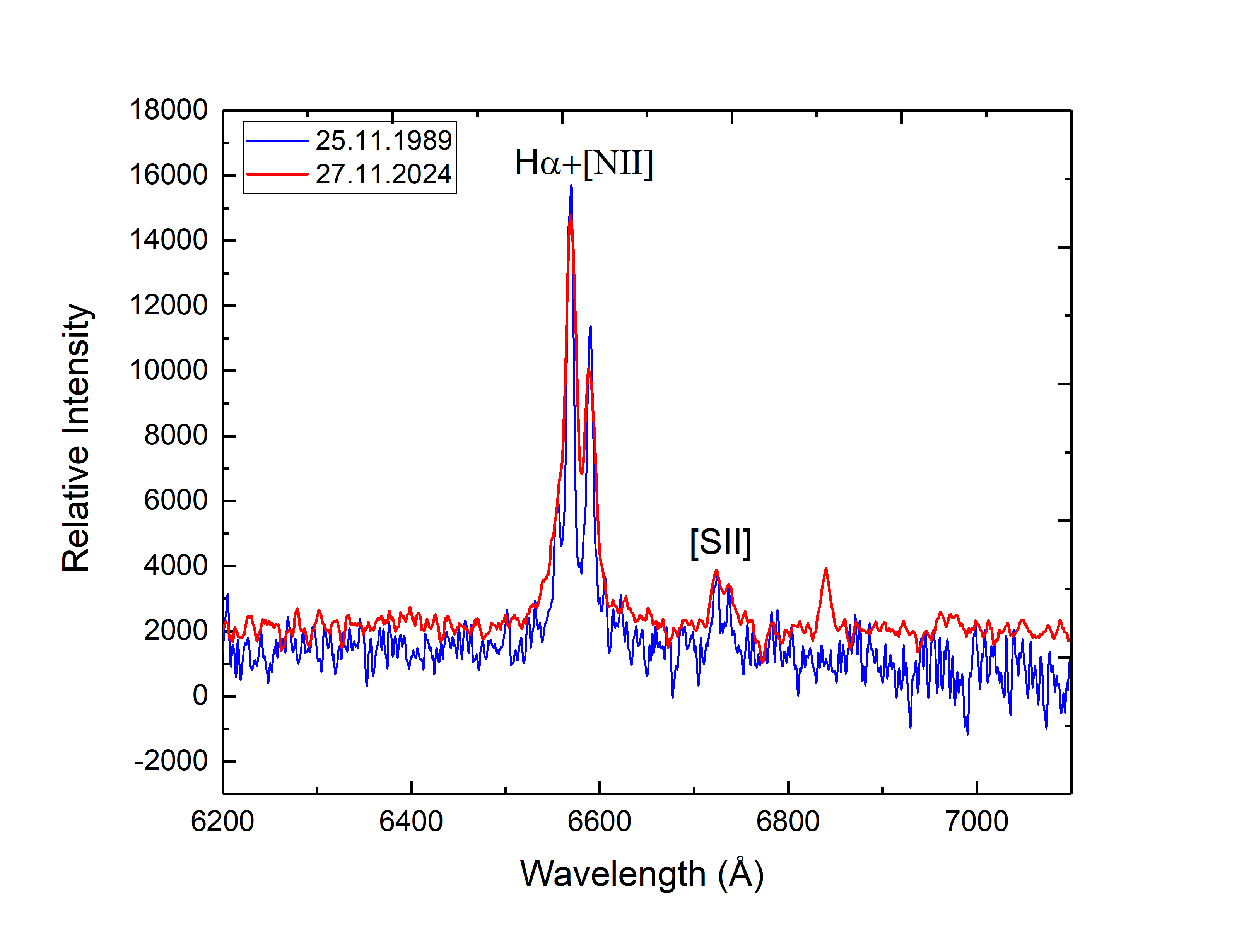}
 \caption{Cross-epoch spectral alignment for Seyfert galaxy NGC 7469: digitized 1989 IIT film spectrum (blue) vs. modern 2024 CCD spectrum (red) in the H$\alpha$+[N\,II] and [S\,II] spectral region.\label{fig:time_domain_case}}
 
\end{figure}

Cross-matching archival holdings with modern observations enables direct multi-epoch analysis. The use case of the approach was presented by \citep{2019ApJS..241...33L}, who combined historical spectroscopic series with modern photometric monitoring to reveal long-term periodic variability in Ark~120. Here, using the Seyfert galaxy NGC~7469 as a \mbox{benchmark \citep{Shomshekova2025}}, we illustrate this capability across a 35-year spectral baseline. By combining an archival spectrum obtained with an image intensifier tube (IIT UM-92, 25~November~1989) and a modern CCD spectrum (27~November~2024) at the 70-cm AZT-8 telescope, we highlight the framework's effectiveness. Standardizing metadata within the \texttt{ivoa.obscore} data model supports the discovery and selection of spectra for comparative analysis. Wavelength calibration and any resampling required for comparison are separate data-processing operations, rather than functions of ObsCore. Figure~\ref{fig:time_domain_case} illustrates the comparison of the H$\alpha$+[N\,II] blend and [S\,II] doublet in the selected spectra.

\begin{figure}[H]
\includegraphics[width=0.9\textwidth]{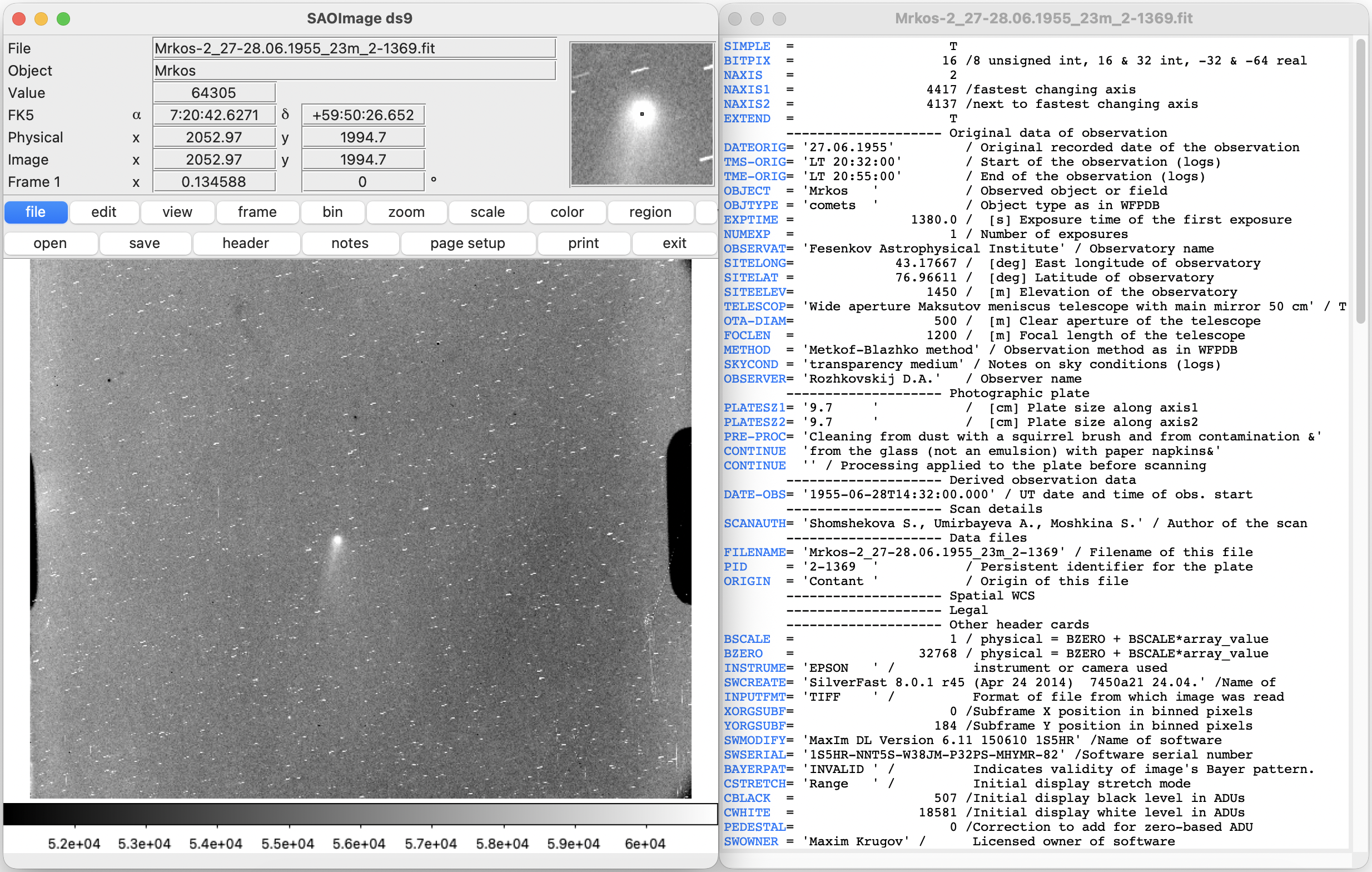}
\caption{Digitized FITS frame of comet C/1955 L1 (Mrkos) displaying the fully populated WCS header within SAOImage~DS9 software (Smithsonian Astrophysical Observatory, Cambridge, MA, USA) \citep{fruscione2026ds9}.
\label{fig:mrkos_ds9}}
\end{figure}

The complete transformation cycle for historical glass archives is illustrated using a digitized photographic plate of comet C/1955 L1 (Mrkos) acquired on 27 June 1955 with ASI-2 telescope (Figure~\ref{fig:mrkos_ds9}). In contrast to the raw physical media inspectable on transmission light (Section~\ref{subsec:data_materials}), the {published digitized FITS product with a WCS solution} undergoes automated grayscale inversion (negative-to-positive) and full header enrichment with calibrated WCS solutions compliant with IVOA plate standards.

\vspace{-7pt} 

\begin{figure}[H]
\includegraphics[width=1\textwidth]{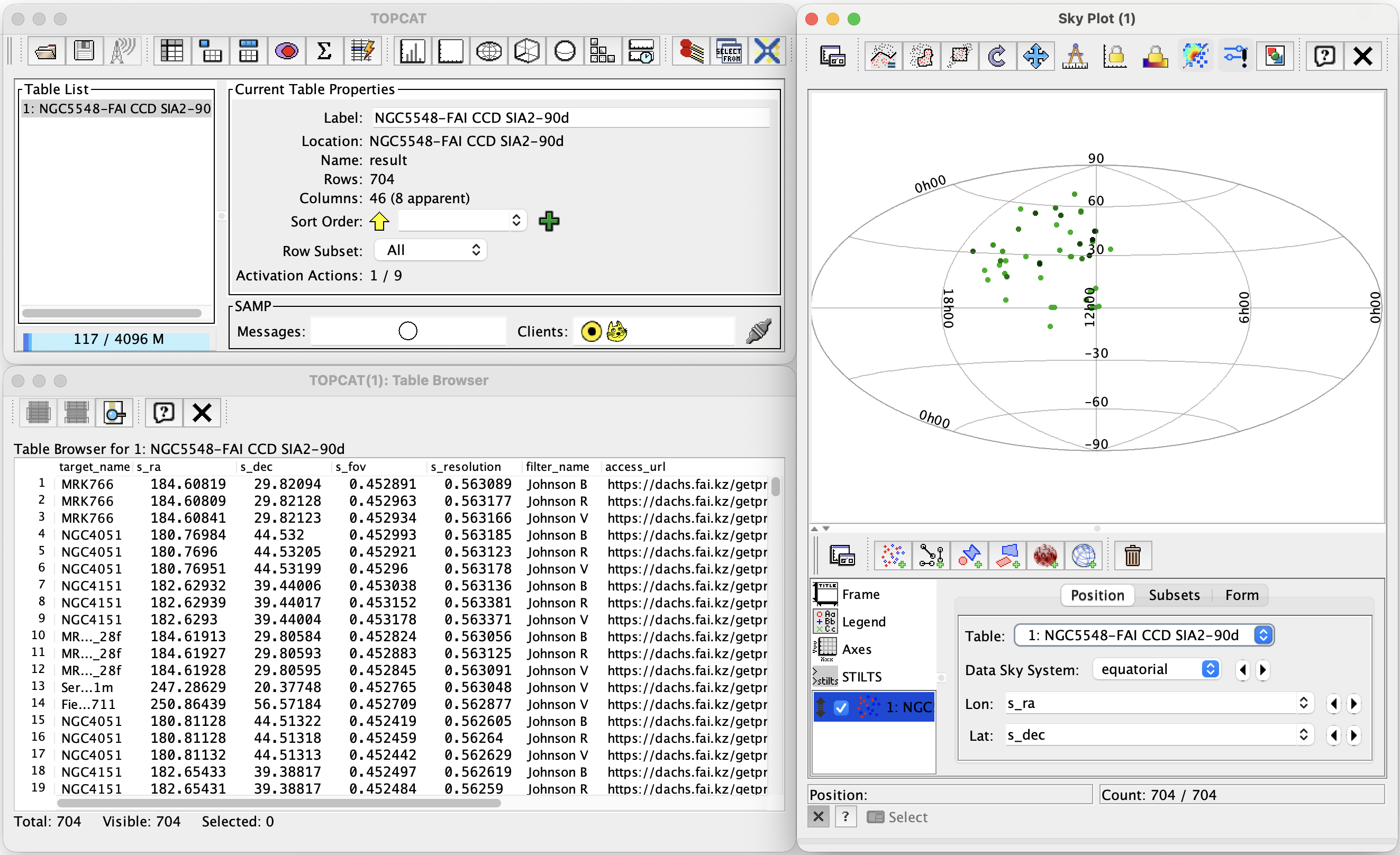}
\caption{Visualization of SIAP service query execution and data discovery for the Zeiss-1000~(East) modern CCD photometric collection within TOPCAT, centered on NGC~5548 with a $90^\circ$ search radius: top-left shows the main application window, bottom-left displays the retrieved ObsCore/SIAP metadata table yielding 704~individual frames, and right presents the spatial footprint of observations on an Aitoff projection.\label{fig:topcat_service}}
\end{figure}

{Registry discovery enables external applications, such as TOPCAT, to locate KazVO services and query their holdings through the corresponding data-access protocols.} {\mbox{Figure~\ref{fig:topcat_service}} shows a SIAP query for the Zeiss-1000 (East) CCD collection, centered on NGC~5548 with a $90^\circ$ search radius. The panels display the TOPCAT interface, the returned image metadata, and the spatial distribution of the retrieved products in an Aitoff projection.}

Thus, the unified representation of heterogeneous KazVO holdings seamlessly integrates archival and current observational data into a scientific workflow, ensuring full readiness for global astronomical toolchains.

\subsection{Technical Enforcement of Data Policy and Embargo Framework} \label{subsec:embargo_enforcement}

The practical enforcement of data embargo policies within KazVO relies on a hybrid payload masking engine coupled with automated access control at the PostgreSQL database and GAVO DaCHS microservice levels.

\subsubsection{Zero-Filled Payload Masking Algorithm}

\begin{figure}[H]
\includegraphics[width=1\textwidth]{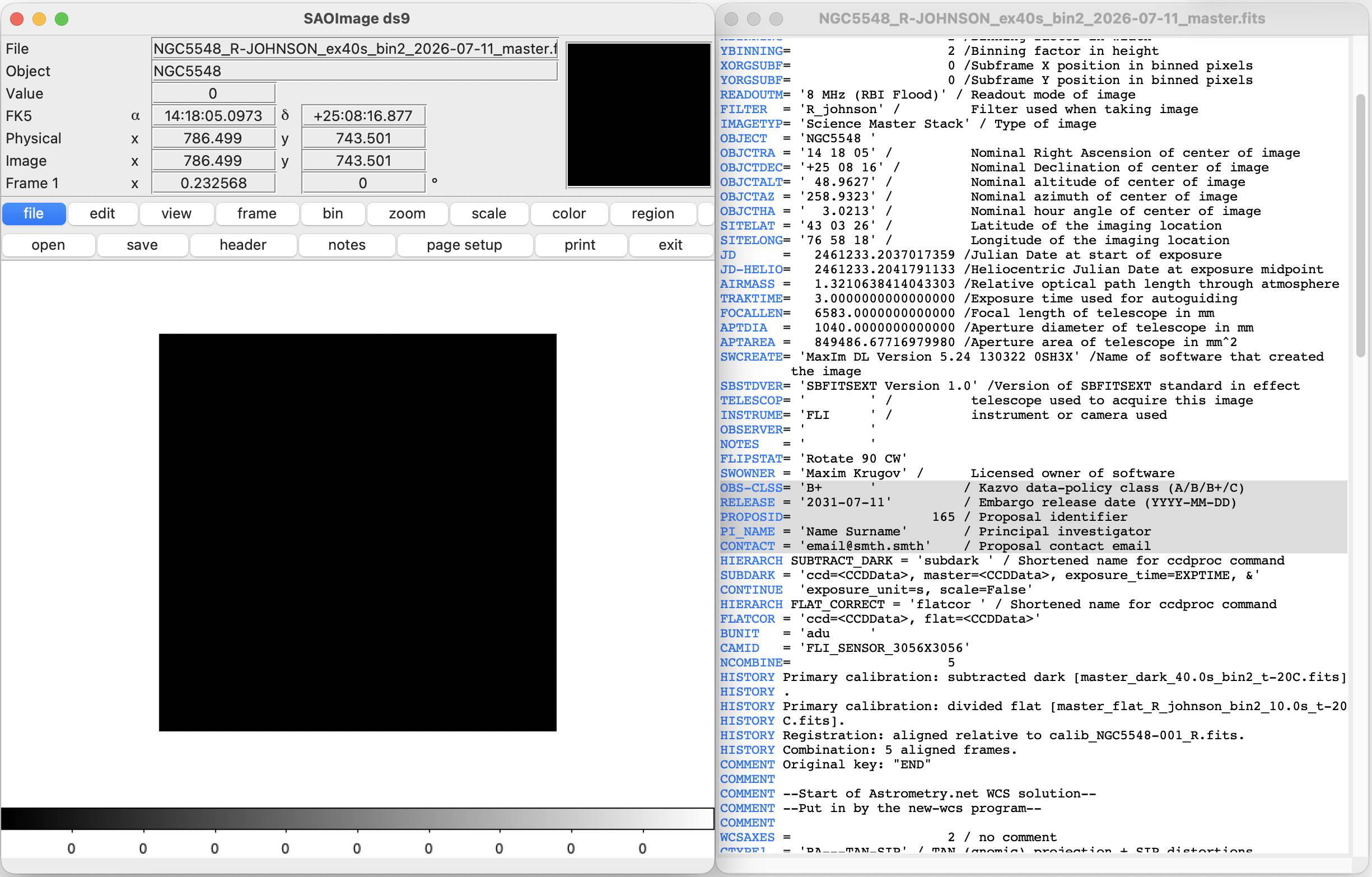}
\caption{{FITS image masking during an active embargo. \textbf{Left}: visualization of the zero-filled image array. \textbf{Right}: the retained primary FITS header, including WCS metadata and highlighted governance keywords.}}
\label{fig:showcase_fts}
\end{figure}

For datasets currently under active embargo (classes A, B, and B+), direct physical access to raw or calibrated binary arrays is restricted. To ensure search transparency and metadata discoverability via the SIAP and SSAP protocols, the data delivery server generates a protected version of the FITS file (``Showcase FITS''):
\begin{enumerate}
    \item \textbf{Header Preservation:} The primary FITS header block is fully retained, including WCS astrometric keys (\texttt{CRVAL}, \texttt{CDELT}, \texttt{SIP}), exposure parameters, and instrument identifiers, all necessary for spatial and temporal coverage validation.
    \item \textbf{Payload Zeroing:} The binary data array of image pixels (\texttt{NAXIS = 2}) or spectral bins (\texttt{NAXIS = 1}) is replaced with an array of zero-valued elements matching the original data type (\texttt{BITPIX}) and dimensions.   {When visualized in astronomical software such as DS9, the masked image displays a uniform zero-filled field, while the primary header remains available for metadata inspection (Figure~\ref{fig:showcase_fts}). The tested masking handler preserves the primary header and does not perform automatic anonymization of observer identifiers.}

\end{enumerate}

The masking mechanism protects the data arrays during embargo; the request-time selection between masked and original content is described and evaluated below.

\subsubsection{Dynamic Routing and Automated Transition to Public}

{The CCD file-delivery service determines access at request time using the embargo date stored in \texttt{dc.products}. The Python handler compares this date with the current server date obtained through \texttt{datetime.date.today()}. Before the release date, the service returns a zero-filled Showcase FITS; on or after that date, it selects the original FITS file without requiring a manual change of access status.}

{We evaluated this release logic on the development host using the handler code extracted directly from the deployed resource descriptor. A synthetic FITS file contained two image HDUs with nonzero 16-bit integer and 32-bit floating-point arrays. The release date remained fixed, while a controlled test clock represented the day before release, the release date itself, and the following day. Database lookup was replaced by a test double returning the same synthetic product and embargo date throughout the test.}

{All three date cases passed: before release, both image arrays were zero-valued; on and after release, the returned file was byte-identical to the original. FITS header blocks were preserved byte-for-byte in all cases, and the source file remained unchanged. An additional check returning the test clock to the pre-release date reproduced the masked output. Separately, two HTTP regression tests against the running development service passed, confirming SIAP metadata retrieval and masked delivery of an embargoed image.}

{These results validate the handler's release-date logic under controlled conditions and the separately tested HTTP operations. They do not constitute an observed release event across an actual date boundary in the running service. The validation script, extracted service handler, recorded controlled-test results, and reproduction instructions are available in the project repository \citep{fai_sort_repo}.}

\subsection{Operational KazVO Data Holdings and Registry Integration}\label{subsec:res_holdings}

Building upon the foundational Data Lake infrastructure and automated reduction pipelines described above, the KazVO framework successfully transforms heterogeneous observational data streams and scientific collections into {a common publication infrastructure based on IVOA standards}. Beyond raw and calibrated observational archives, the ecosystem incorporates multi-wavelength astronomical catalogs, theoretical simulation grids, continuous space weather monitoring series, and dedicated in-house web utilities.

{Published resources are described through GAVO DaCHS resource descriptors using metadata appropriate to each resource type. ObsCore calibration levels apply to the observational products represented in the ObsCore view. Standalone catalogs, observational logs, and theoretical simulation grids are described through their own table schemas; no ObsCore \texttt{calib\_level} is reported for these resources in Table~\ref{tab:kazvo_holdings}.} Access to the infrastructure is architected in a hybrid mode: human-centric interactive querying and graphical inspection are provided via the primary web portal (\url{http://vo.fai.kz}, accessed on 3~November~2026), whereas programmatic, M2M workflows and integration with external client tools, e.g., PyVO, TOPCAT, Aladin, are driven by standardized IVOA protocols via the primary TAP service endpoint assigned the root IVOID: \texttt{ivo://fai.kz/tap}. Table~\ref{tab:kazvo_holdings} presents a consolidated overview of data collections, access protocols, processing levels, and record volumes across all published KazVO resources.

\begin{table}[H]
\caption{{Consolidated architecture and quantitative metrics of published data holdings, analytical catalogs, and services in the KazVO node (snapshot of the production GAVO DaCHS registry, PostgreSQL backend, and filesystem on 28 October 2026).}}
\label{tab:kazvo_holdings}
\begin{adjustwidth}{-\extralength}{0cm}
\footnotesize
\setlength{\tabcolsep}{2pt}
\begin{tabular*}{\fulllength}{@{\extracolsep{\fill}}lllccccc@{}}
\toprule
\textbf{KazVO Service} & \textbf{Data Type} & \textbf{Access} & \textbf{ObsCore} & \textbf{DB} & \textbf{DB} & \textbf{Data} & \textbf{File} \\
{} & {} & \textbf{Protocol} & \textbf{Level \textsuperscript{1}} & \textbf{Rows \textsuperscript{2}} & \textbf{Size \textsuperscript{3}} & \textbf{Files \textsuperscript{4}} & \textbf{Volume \textsuperscript{5}} \\
\midrule
\multicolumn{8}{@{}l}{\textit{1. Observational FITS Archives}} \\
ASI-2 Glass Archive & FITS Image & SIAP & 1 & 1713 & 1.78 MB & 1713 & 57.00 GB \\
Schmidt Camera Glass Archive & FITS Image & SIAP & 1 & 1316 \textsuperscript{a} & 1.11 MB & 1316 \textsuperscript{a} & 48.81 GB \textsuperscript{a} \\
AZT-8 Planetary Nebulae Archive & Slit Spectrum & SSAP & 2 & 672 & 663.55 KB & 672 & 10.51 MB \\
AZT-8 AGN Spectral Archive & Slit Spectrum & SSAP & 2 & 1055 & 851.97 KB & 1055 & 19.33 MB \\
Zeiss-1000 CCD Photom. Archive & FITS Image & SIAP & 3 & 5219 & 6.59 MB & 5198 \textsuperscript{b} & 136.63 GB \textsuperscript{b} \\
FAI Echelle Spectra Observations & Echelle Sp. & SSAP & 2 & 8 & 73.73 KB & 8 & 1.52 MB \\
TCO Optical Echelle Spec. Archive & Echelle Sp. & SSAP/SODA & 3 & 837 & 1.07 MB & 837 & 137.15 MB \\
TCO Echelle Sp. (Hot Supergiants) & Echelle Sp. & SSAP/SODA & 3 & 512 & 647.17 KB & 512 & 81.66 MB \\
\midrule
\multicolumn{8}{@{}l}{\textit{2. Astronomical Catalogs \& Processed Tables}} \\
ML-Enhanced GCVS (Gaia/TESS) & Value-added Cat. & TAP/SCS & N/A & 55,376 & 14.06 MB & 1 & 17.74 MB \\
Wolf-Rayet Stars (WC/WO) & Spectral Line Tab. & TAP & N/A & 100 & 49.15 KB & 1 & 24.68 KB \\
Wolf-Rayet Stars (WN) & Spectral Line Tab. & TAP & N/A & 159 & 57.34 KB & 1 & 35.37 KB \\
Planetary Nebulae Abs. Spec. & Calculated Params & TAP & N/A & 166 & 98.30 KB & 1 & 28.57 KB \\
Spectrophotom. Standards & Flux Standard Cat. & TAP & N/A & 1273 & 1.12 MB & 1 & 557.51 KB \\
Galactic X-ray Pulsars Catalog & HMXB Cat. & TAP/SCS & N/A & 81 & 57.34 KB & 1 & 36.86 KB \\
FAI Asteroid Observation Log & Observational Log & TAP & N/A & 10,296 & 1.37 MB & 1 & 27.15 KB \\
\midrule
\multicolumn{8}{@{}l}{\textit{3. Space Weather Monitoring Series}} \\
Alma-Ata Geomagnetic Obs. & Magnetometer Series & TAP & 2 & 2,226,240 & 356.28 MB & 1549 & 112.10 MB \\
Solar Radio Emission (``Orbita'') & Radio Flux Series & TAP & 1 & 5,971,775 & 950.58 MB & 1204 & 184.83 MB \\
Alma-Ata Neutron Monitor & Cosmic Ray Intensity & TAP & 1 & 2,231,558 & 355.93 MB & 1383 & 71.44 MB \\
\midrule
\multicolumn{8}{@{}l}{\textit{4. Theoretical Astrophysics \& Simulations}} \\
MESA Grid: Binary Evolution & Evolutionary Grid & TAP & N/A & 1414 & 1.20 MB & 15 \textsuperscript{c} & 7.04 MB \textsuperscript{c} \\
\midrule
\multicolumn{8}{@{}l}{\textit{5. Interactive Web Services \& Software}} \\
ArXSP & Spectrum processing & Desktop & N/A & --- & --- & --- & --- \\
Wavelength Calibration Tool & DTW/RANSAC tool & Web & N/A & --- & --- & --- & --- \\
Astronomical Calendar & Twilight calculator & Web & N/A & --- & --- & --- & --- \\
Order Computations & Computation requests & Web & N/A & --- & --- & --- & --- \\
Order Observations & Observation requests & Web & N/A & --- & --- & --- & --- \\
\midrule
\textbf{Total Listed} & \multicolumn{2}{l}{\textbf{19 data collections + 5 tools/services}} & & \textbf{10,509,770} & \textbf{1.69 GB} & \textbf{15,469} & \textbf{243.08 GB} \\
\bottomrule
\end{tabular*}
\end{adjustwidth}
\noindent{\footnotesize{\textsuperscript{1}~Integer calibration levels (\texttt{calib\_level}) are assigned in accordance with IVOA ObsCore v1.1 \citep{ivoa.obscore2017} categories (1: digitized plates and instrumental detector streams in standard formats; 2: calibrated science-ready spectra and geomagnetic series; 3: enhanced products, including co-added CCD master stacks and merged multi-order echelle spectra). These values reflect physical data processing status rather than validation or service maturity. For space weather detector time-series queried via dedicated TAP tables, the levels indicate equivalent instrumental calibration states without implying inclusion in the \texttt{ivoa.obscore} view. In the current KazVO implementation, the listed standalone catalogs, observational logs, and simulation grids are published through dedicated TAP tables rather than as ObsCore observation products. Their calibration level is therefore marked N/A in this table; N/A also applies to software tools. \textsuperscript{2}~Exact row counts (\texttt{COUNT(*)}) stored across physical PostgreSQL tables belonging to the resource descriptor (excluding ordinary SQL views to avoid double counting). \textsuperscript{3}~Physical PostgreSQL storage footprint (\texttt{pg\_total\_relation\_size}), encompassing table heap, TOAST storage, and B-tree/GiST spatial indices ($1\text{ MB} = 10^6\text{ bytes}$, $1\text{ KB} = 10^3\text{ bytes}$). \textsuperscript{4}~Distinct data files counted directly on disk, with symbolic and hard-link aliases deduplicated by filesystem device and inode. For observational archives, the count covers registered downloadable products without an active embargo; for catalogs and space-weather series, it covers the source files used for table ingestion (FITS, CSV, or SQLite). These source files are not necessarily exposed as individual VO download products. MESA includes model files and the consolidated ingestion table. Administrative files, previews, and auxiliary processing copies are excluded. \textsuperscript{5}~Sum of the logical sizes of the files counted in the preceding column, measured using filesystem metadata; this is not allocated NAS storage or a measurement of the entire Data Lake. Displayed sizes are rounded; exact byte totals are retained in the audit. Decimal units are used ($1\text{ GB}=10^9\text{ bytes}$, $1\text{ MB}=10^6\text{ bytes}$, $1\text{ KB}=10^3\text{ bytes}$). \textsuperscript{a}~Includes 1,216 science target plates in the primary query view plus 100 associated calibration rows/files. 
\textsuperscript{b}~Excludes 21 embargoed CCD products (700,850,880 bytes). Across the observational archives and the 14 downloadable MESA models, there are 11,325 non-embargoed registered file products (242,691,888,961 bytes); the larger total in this table additionally includes catalog and monitoring ingestion files and the consolidated MESA table.
\textsuperscript{c}~Fourteen model FITS files (5,967,360 bytes) and one consolidated FITS table used for ingestion (1,068,480 bytes), totaling 15 files and 7,035,840 bytes. The corresponding PostgreSQL resource has one physical table. The consolidated file is a derived representation, not an additional independent model.}}
\end{table}

\vspace{6pt} 

{Table~\ref{tab:kazvo_holdings} lists 19 data collections and five research tools/services. Their access interfaces are specified individually; collection counts do not represent counts of independent IVOA endpoints. The DaCHS backend provides the protocol implementation and diagnostic tools used to check resource configurations and service responses.} 

\section{Discussion}\label{sec:disc}

The implementation of the KazVO infrastructure demonstrates that deploying {a node implementing established IVOA interfaces} does not require building a software ecosystem from scratch, but rather relies on the systematic harmonization of heterogeneous archives using standardized protocols and containerized microservices. KazVO addresses the challenges of data isolation and scientific heritage preservation while ensuring the integration of local observational facilities into the global infrastructure.

The ``Scientific Showcase'' mechanism implemented in KazVO is based on practices widely adopted across leading international data centers for balancing open science interests with proprietary rights. The automated preservation of primary headers and spatial WCS metadata, combined with the dynamic masking of binary payloads during the active embargo period, resolves the conflict between observer priorities and open-access standards. This enables the global community to verify spatial and temporal coverage via the SIAP and SSAP protocols without premature disclosure of confidential data.

{The historical Glass Library comprises approximately 20,000 physical plates and films. Its digitized products are being published in stages through KazVO services; the current holdings listed in Table~\ref{tab:kazvo_holdings} include 4,756 historical image and spectral file products, including associated calibration files. The size of the physical collection should therefore be distinguished from the extent of the published release.} Combined with real-time space weather monitoring data streams and modern photometric pipelines at ATO and TShAO, KazVO operates as a multi-domain Data Lake rather than a single-instrument archive. Nevertheless, certain limitations remain and are actively being addressed: the automation of spectroscopic reduction pipelines is currently under development, and the process of routine automated data ingestion from observational sites is at the implementation stage. Future work will focus on transitioning to operational observational data flows as data accumulates, as well as deepening integration within the AstroHub ecosystem.

\section{Conclusions} \label{sec:concl}

This paper presents the architectural principles, data lifecycle management, and practical implementation of the Kazakhstani National Virtual Observatory (KazVO). The primary findings of this study are summarized as follows:

\begin{enumerate}
    \item \textbf{Infrastructural Consolidation:} A centralized Data Lake has been established on the Synology HA NAS and Proxmox VE infrastructure. The broader inventory summarized in Table~\ref{tab:datalake_summary} accounts for approximately 9.70~TB of listed holdings, including raw observations, calibration files, intermediate products, logs, and processed data.
    \item \textbf{Interoperability and Service Publication:} Table~\ref{tab:kazvo_holdings} lists 19 data collections and five tools/services. The production audit of 28 October 2026 measured 10,509,770 database rows occupying 1.69~GB, including indices, and 15,469 associated data files totaling 243.08~GB. The file total includes catalog and monitoring ingestion files and the consolidated MESA table; it is not exclusively a count of publicly downloadable products. The non-embargoed registered download products comprise 11,325 files totaling 242.69~GB.
    \item \textbf{Embargo Policy Implementation:} The ``Scientific Showcase'' module preserves searchable observational metadata while restricting access to the original data arrays during proprietary embargo periods. Controlled tests of the deployed CCD delivery handler confirmed masked output before the release date and byte-identical original output on and after that date. An operational release event across an actual date boundary has not yet been demonstrated.
    \item \textbf{Archive Preservation and Integration:} A comprehensive digitization, astrometric calibration, and metadata standardization workflow for the historical photographic plate archive (1950--1997) has been established. To date, {4756 historical plate images and spectral products} have been successfully {made available through registered VO services}, while the remaining digitized legacy holdings in the Data Lake undergo staged astrometric and spectrophotometric reduction.
\end{enumerate}

The operational scope of the current implementation comprises the published VO services, targeted photometric processing, and scheduled, batch-mode ingestion. Replication has been validated on pilot datasets but has not yet been demonstrated as a routine workflow across the observing sites. Automated reduction of modern spectroscopic streams remains under development.

Reaching routine operation across the complete data lifecycle will require deployment and sustained validation of replication at the observing sites, together with integration and validation of the spectroscopic processing pipelines.

KazVO provides a common infrastructure for publishing historical
and modern astronomical holdings through standardized IVOA interfaces. The present deployment establishes the foundation for extending the operational services towards a more automated data lifecycle as the remaining ingestion and processing components reach routine operation.

\vspace{6pt}
\authorcontributions{Conceptualization, I.I., D.Y., Y.A., M.M., A.S., S.S., V.K., C.O., G.A., R.V. and R.K.; methodology, I.I., D.Y., M.M., A.S., S.S., V.K., I.R. and G.A.; software, I.I., M.M., V.K., A.U., A.G.; validation, I.I., S.S., A.U. and L.A.; formal analysis, I.I., M.M., S.S., A.U., L.A., N.V.; investigation, S.S., A.U., L.A., M.K., N.V., D.A. and I.R.; resources, D.Y., M.K. and C.O.; data curation, I.I., M.M., S.S., A.U. and L.A.; writing---original draft preparation, I.I., D.Y., Y.A., A.S., S.S., V.K. ; writing---review and editing, I.I., Y.A., D.K. and G.S.; visualization, I.I., Y.A. and S.S.; supervision, D.Y.; project administration, D.Y. and C.O.; funding acquisition, I.I., D.Y., Y.A., M.M., A.S., S.S., V.K., C.O., G.A., R.V. and R.K.
All authors have read and agreed to the published version of the manuscript.
}

\funding{This research has been funded by the Committee of Science of the Ministry of Science and Higher Education of the Republic of Kazakhstan (Grant No. BR24992807).}

\institutionalreview{Not applicable.}

\informedconsent{Not applicable.}

\dataavailability{The astronomical datasets presented in this study comprise both heritage photographic collections and modern observational streams. The digitized legacy archives and published target datasets are publicly accessible via the Kazakhstani Virtual Observatory (KazVO) TAP/SIAP services at \url{http://vo.fai.kz} accessed on 3~November~2026, and can be queried directly through IVOA-compliant clients (e.g., TOPCAT, Aladin). {Access to modern CCD products follows the applicable collection policy. For products governed by a finite embargo date, the file-delivery handler selects masked or original content according to that date at request time. This behavior has been verified in controlled tests using a simulated clock; an operational transition across an actual release date has not yet been demonstrated. Products subject to indefinite restrictions are not covered by automatic release.}}

\acknowledgments{The authors express their sincere gratitude to Markus Demleitner for his invaluable consultations, endless patience, and continued support. We are deeply indebted to all observatory staff and observers for collecting the observational data, even if deciphering their handwritten logbooks occasionally tested our patience. We also thank the entire staff of the Fesenkov Astrophysical Institute for their contributions to the institute's research activities. Finally, we extend our heartfelt appreciation to the entire IVOA community for stimulating InterOps, engaging discussions, and fostering an exceptionally welcoming and collaborative environment. During the preparation of this manuscript, the authors used Gemini (Gemini 3.5 Flash-Lite) for assistance in language polishing, stylistic refinement, and translation. The authors have reviewed and edited the output and take full responsibility for the content of this publication.
}

\conflictsofinterest{The authors declare no conflicts of interest. The funders had no role in the design of the study; in the collection, analyses, or interpretation of data; in the writing of the manuscript; or in the decision to publish the results.} 



\abbreviations{Abbreviations}{%
The following abbreviations are used in this manuscript:\\

\noindent 
\begin{tabular}{@{}m{2cm}<{\raggedright}m{10cm}<{\raggedright}}
ADQL & Astronomical Data Query Language\\
API & Application Programming Interface\\
ArVO & Armenian Virtual Observatory\\
AstroHub & Astronomical Hub\\
ATO & Assy-Turgen Observatory\\
CC BY & Creative Commons Attribution\\
CDS & Strasbourg Astronomical Data Center\\
CPU & Central Processing Unit\\
CUDA & Compute Unified Device Architecture\\
DaCHS & Data Center Helper Suite\\
DOI & Digital Object Identifier\\
ESO & European Southern Observatory \\
FAI & Fesenkov Astrophysical Institute\\
FAIR & Findable, Accessible, Interoperable, Reusable\\
FFDB & FITS Header Database Service\\
FITS & Flexible Image Transport System\\

GAVO & German Astrophysical Virtual Observatory\\

GPU & Graphics Processing Unit\\
HA & High Availability\\

\end{tabular}

\noindent
\begin{tabular}{@{}m{2cm}<{\raggedright}m{10cm}<{\raggedright}}
HST & Hubble Space Telescope\\
IIT & Image Intensifier Tube\\
IRAF & Image Reduction and Analysis Facility\\
IVOA & International Virtual Observatory Alliance\\
IVOID & International Virtual Observatory Identifier\\
JSON & JavaScript Object Notation\\
KazVO & Kazakhstani National Virtual Observatory\\
LT & Local time\\
LST & Local sidereal time\\
LTS & Long-Term Support\\
LXC & Linux Containers\\
M2M & {Machine-to-Machine}\\
ML & Machine Learning\\
MPI & Message Passing Interface\\
NAS & Network-Attached Storage\\
NRAO & National Radio Astronomy Observatory\\
OS & Operating System\\
RofR & {Registry of Registries}\\
RVO & Russian Virtual Observatory\\
SAMP & Simple Application Messaging Protocol\\
SIAP & Simple Image Access Protocol\\
SLURM & Simple Linux Utility for Resource Management\\
SODA & Server-side Operations for Data Access\\
SSAP & Simple Spectral Access Protocol\\
TAP & Table Access Protocol\\
TAPRegExt & Table Access Protocol Registry Extension\\
TFLOPS & Trillion Floating-Point Operations Per Second\\
TIFF & Tagged Image File Format\\
TOPCAT & Tool for OPerations on Catalogues And Tables\\
TShAO & Tien-Shan Astronomical Observatory\\
URI & Uniform Resource Identifier\\
UTC & Coordinated Universal Time\\
VOSI & IVOA Support Interfaces\\
WCS & World Coordinate System\\
\end{tabular}
}

\begin{adjustwidth}{-\extralength}{0cm}
\reftitle{References}


\PublishersNote{}
\end{adjustwidth}

\begin{thebibliography}{999}

\bibitem{unesco2021openscience}
UNESCO. {\em UNESCO Recommendation on Open Science}; UNESCO: Paris, France, 2021. \url{https://doi.org/10.54677/MNMH8546}.

\bibitem{reitze2024}
Reitze, D. The Evolution of Astrophysics towards Big Science: Insights from the Innovation Landscape. In {\em Big Science, Innovation, and Societal Contributions: The Organisations and Collaborations in Big Science Experiments}; Liyanage, S., Nordberg, M., Streit-Bianchi, M., Eds.; Oxford University Press: Oxford, UK, 2024; pp. 185--219. \url{https://doi.org/10.1093/oso/9780198881193.003.0009}.

\bibitem{zhang2015}
Zhang, Y.; Zhao, Y. Astronomy in the Big Data Era. {\em Data Sci. J.} {\bf 2015}, {\em 14}, 11. \url{https://doi.org/10.5334/dsj-2015-011}.

\bibitem{aimuratov2025}
Aimuratov, Y.; Kim, V.; Serebryanskiy, A.; Yurin, D.; Krugov, M.; Akniyazov, C.; Shomshekova, S.; Makukov, M.; Aimanova, G.; Valiullin, R.; et al. The Astronomical Hub: A Unified Ecosystem for Modern Astronomical Research. {\em Galaxies} {\bf 2025}, {\em 13}, 99. \url{https://doi.org/10.3390/galaxies13050099}.

\bibitem{astro2020}
National Academies of Sciences, Engineering, and Medicine. \textit{Pathways to Discovery in Astronomy and Astrophysics for the 2020s}; The National Academies Press: Washington, DC, USA, 2021; pp. 128--156. \url{https://doi.org/10.17226/26141}. 

\bibitem{norris2006}
Norris, R.; Andernach, H.; Genova, F.; Griffin, E.; Hanisch, R.; Kembhavi, A.; Kennicutt, R.; Richards, A. Astronomical Data Management. {\em arXiv} {\bf 2006}, arXiv:astro-ph/0612628.

\bibitem{norris2007}
Norris, R. How to Make the Dream Come True: The Astronomers---Data Manifesto. {\em Data Sci. J.} {\bf 2007}, {\em 6}, 116--124. \url{https://doi.org/10.48550/arXiv.astro-ph/0701361}.

\bibitem{desouza2025}
de Souza, R.S.; Ishida, E.E.O.; Krone-Martins, A. A Brief History of Inference in Astronomy. {\em arXiv} {\bf 2025}, arXiv:2510.17433. \url{https://doi.org/10.48550/arXiv.2510.17433}.

\bibitem{huppenkothen2023}
Huppenkothen, D.; Ntampaka, M.; Ho, M.; Fouesneau, M.; Nord, B.; Peek, J.E.G.; Walmsley, M.; Wu, J.F.; Avestruz, C.; Buck, T.; et al. Constructing Impactful Machine Learning Research for Astronomy: Best Practices for Researchers and Reviewers. {\em arXiv} {\bf 2023}, arXiv:2310.12528. \url{https://doi.org/10.48550/arXiv.2310.12528}.

\bibitem{brescia2024}
Brescia, M.; Angora, G. Strengthening leverage of Astroinformatics in inter-disciplinary Science. {\em arXiv} {\bf 2024}, arXiv:2409.03425. \url{https://doi.org/10.48550/arXiv.2409.03425}.

\bibitem{erard2025sf2a}
Erard, S.; Le Sidaner, P.; Boisson, C.; Meliani, Z. Paris Astronomical Data Centre: 20 Years and Counting. In Proceedings of the SF2A-2025: Semaine de l'Astrophysique Française, Toulouse, France, 1--4 July 2025; pp. 23--25.

\bibitem{wilkinson2016fair}
Wilkinson, M.D.; Dumontier, M.; Aalbersberg, I.J.; Appleton, G.; Axton, M.; Baak, A.; Blomberg, N.; Boiten, J.W.; da Silva Santos, L.B.; Bourne, P.E.; et al. The FAIR Guiding Principles for Scientific Data Management and Stewardship. {\em Sci. Data} {\bf 2016}, {\em 3}, 160018. \url{https://doi.org/10.1038/sdata.2016.18}.

\bibitem{landais2024adass}
Landais, G.; Ocvirk, P.; Allen, M.; Bonnarel, F.; Perret, E.; Vannier, P.; Brouty, M.; Fix, C.; Monari, G. Build FAIR Workflow for Astronomical Catalogues. In {\em Astronomical Data Analysis Software and Systems XXXI}; Hugo, B.V., Van Rooyen, R., Smirnov, O.M., Eds.; ASP Conference Series; Astronomical Society of the Pacific: San Francisco, CA, USA, 2024; Volume 535, p. 295.

\bibitem{otoole2024adass}
O'Toole, S.; Tocknell, J. FAIR Standards for Astronomical Data. In {\em Astronomical Data Analysis Software and Systems XXXI}; Hugo, B.V., Van Rooyen, R., Smirnov, O.M., Eds.; ASP Conference Series; Astronomical Society of the Pacific: San Francisco, CA, USA, 2024; Volume 535, p. 265.

\bibitem{dowler2021arch}
Dowler, P.; Evans, J.; Arviset, C.; Gaudet, S.; Technical Coordination Group. IVOA Architecture (Version 2.0). {IVOA Endorsed Note}. { 2021}. Available online: \url{https://wiki.ivoa.net/twiki/bin/view/IVOA/IvoaTCG} (accessed on 23~September~2026).

\bibitem{dowler2019tap}
Dowler, P.; Rixon, G.; Tody, D.; Demleitner, M. Table Access Protocol (Version 1.1). IVOA Recommendation {2019}. Available online: \url{https://www.ivoa.net/documents/TAP/} (accessed on 23~September~2026).

\bibitem{dowler2015sia}
Dowler, P.; Tody, D.; Bonnarel, F. IVOA Simple Image Access (Version 2.0). {IVOA Recommendation}. {2015}. Available online: \url{https://ui.adsabs.harvard.edu/abs/2015ivoa.spec.1223D/abstract} (accessed on 3~November~2026). 


\bibitem{tody2012ssap}
Tody, D.; Dolensky, M.; McDowell, J.; Bonnarel, F.; Budavari, T.; Busko, I.; Micol, A.; Osuna, P.; Salgado, J.; Skoda, P.; et al. Simple Spectral Access Protocol (Version 1.1). { IVOA Recommendation}. { 2012}. Available online: \url{https://ui.adsabs.harvard.edu/abs/2012ivoa.spec.0210T/abstract} (accessed on 3~November~2026).

\bibitem{landais2026ivoa}
Landais, G.; Raugh, A.C.; Muench, A.; Cecconi, B.; D'Abrusco, R.; Henneken, E.; Jenness, T. Best Practices for the Creation of and Metadata for Digital Object Identifiers in Astronomy Archives (Version 1.0). { IVOA Note}. { 2026}. Available online: \url{http://www.ivoa.net/documents/} (accessed on 23~September~2026).

\bibitem{romaniello2023eso}
Romaniello, M.; Arnaboldi, M.; Barbieri, M.; Delmotte, N.; Dobrzycki, A.; Fourniol, N.; Freudling, W.; Grave, J.; Mascetti, L.; Micol, A.; et al. The ESO Science Archive Facility: Status, Impact, and Prospects. {\em Messenger} {\bf 2023}, {\em 191}, 29--33. \url{https://doi.org/10.18727/0722-6691/5338}.

\bibitem{lesteven2025lisa}
Lesteven, S. The CDS in the Landscape of the Open Science. In Proceedings of the LISA 10---Library and Information Services in Astronomy, Santiago, Chile, 3--7~November~2025; p. 37. \url{https://doi.org/10.5281/zenodo.17749621}.

\bibitem{li2017}
Li, C.; Cui, C.; He, B.; Fan, D.; Wang, J.; Li, S.; Mi, L.; Wan, W.; Chen, J.; Zhang, H.; et al. The Design and Application of Astronomy Data Lake in China-VO. In {\em Astronomical Data Analysis Software and Systems XXV}; Lorente, N.P.F., Shortridge, K., Wayth, R., Eds.; ASP Conference Series; Astronomical Society of the Pacific: San Francisco, CA, USA, 2017; Volume 512, p. 157.

\bibitem{dluhznevskaya2018}
Dluhznevskaya, O.B.; Malkov, O.Y. Russian Virtual Observatory. In {\em Stars and Satellites, Proceedings of the Memorial Conference Devoted to A.G. Masevich 100th Anniversary}; Shustov, B.M., Wiebe, D.S., Eds.; INASAN: Moscow, Russia, 2018; pp. 391--396. \url{https://doi.org/10.26087/INASAN.2018.2.2.063}.

\bibitem{mickaelian2025}
Mickaelian, A.M.; Mikayelyan, G.A.; Abrahamyan, H.V.; Astsatryan, H.V.; Knyazyan, A.V. Recent activities of the Armenian Virtual Observatory (ArVO). {\em Commun. Byurakan Astrophys. Obs.} {\bf 2025}, {\em 72}, 3--11. \url{https://doi.org/10.52526/25792776-25.72.1-3}.

\bibitem{hudec2018}
Hudec, R. Astrophysics with digitized astronomical plate archives. {\em Astron. Nachr.} {\bf 2018}, {\em 339}, 408--411. \url{https://doi.org/10.1002/asna.201813515}.

\bibitem{hudec2019}
Hudec, R. Astronomical photographic data archives: Recent status. {\em Astron. Nachr.} {\bf 2019}, {\em 340}, 690--697. \url{https://doi.org/10.1002/asna.201913676}.

\bibitem{wenjing2007}
Wen-Jing, J.; Zheng-Hong, T.; Shu-He, W.; Tian, K.; Chen, L.; Zhao, Y.; Jiang, S.-Y. Review on IAU work for preservation and digitization of astronomical photographic plates and suggestions of plates digitization in China. {\em Prog. Astron.} {\bf 2007}, {\em 25}, 1--12.

\bibitem{rovetto2016}
Rovetto, R.J.; Kelso, T.S. Preliminaries of a Space Situational Awareness Ontology. {\em arXiv} {\bf 2016}, arXiv:1606.01924. \url{https://doi.org/10.48550/arXiv.1606.01924}.

\bibitem{nebot2019}
Nebot, A.; Allen, M.; Fernique, P.; Baumann, M.; Boch, T.; Bot, C.; Derriere, S.; Genova, F.; Lutz, K.; Morris, D. Exploring Time Domain Multi-Messenger Astronomy through the Virtual Observatory. {\em PoS} {\bf 2019}, {\em 357}, 056. \url{https://doi.org/10.22323/1.357.0056}.

\bibitem{denisyuk2003}
Denisyuk, E.K. Spectrograph for Faint Objects: The Device and the Main Results of Observations. {\em Astron. Astrophys. Trans.} {\bf 2003}, {\em 22}, 175--180. \url{https://doi.org/10.1080/1055679031000084795a}.

\bibitem{ivoa.obscore2017}
Louys, M.; Tody, D.; Dowler, P.; Durand, D.; Michel, L.; Bonnarel, F.; Micol, A.; IVOA Data Model Working Group. Observation Data Model Core Components, Its Implementation in the Table Access Protocol Version 1.1. { IVOA Recommendation}, 9 May 2017. Available online:
\url{https://ui.adsabs.harvard.edu/abs/2017ivoa.spec.0509L/abstract} (accessed on 3~November~2026).

\bibitem{demleitner2014dachs}
Demleitner, M.; Neves, M.C.; Rothmaier, F.; Wambsganss, J. Virtual Observatory Publishing with DaCHS. {\em Astron. Comput.} {\bf 2014}, {\em 7}, 27--36. \url{https://doi.org/10.1016/j.ascom.2014.08.003}.

\bibitem{postgresql15}
The PostgreSQL Global Development Group.
PostgreSQL 15 Documentation.
Available online: \url{https://www.postgresql.org/docs/15/}
(accessed on 23~September~2026).



\bibitem{plante2007rofr}
Plante, R. The Registry of Registries (Version 1.00). { IVOA Note}. { 2007}. Available online:
\url{https://ui.adsabs.harvard.edu/abs/2007ivoa.rept.0628P/abstract} (accessed on 3~November~2026).

\bibitem{git_kazvo_inputs}
Fesenkov Astrophysical Institute. Resource Descriptors of KazVO Services. Available online: \url{https://github.com/fai-kz/kazvo-inputs} (accessed on 23~September~2026).

\bibitem{shomshekova2022na}
Shomshekova, S.; Izmailova, I.; Umirbayeva, A.; Omarov, C. A method for digitization of archival astroplates of the Fesenkov Astrophysical Institute. {\em New Astron.} {\bf 2022}, {\em 97}, 101881. \url{https://doi.org/10.1016/j.newast.2022.101881}.

\bibitem{shomshekova2022comets}
Shomshekova, S.A.; Izmailova, I.M.; Moshkina, S.G.; Umirbayeva, A.Z. Digitization of photometric astronegatives of comets from the V.G. Fesenkov Astrophysical Institute. {\em Rep. Natl. Acad. Sci. Repub. Kazakhstan} {\bf 2022}, {\em 1}, 137--143. \url{https://doi.org/10.32014/2022.2518-1483.143}.

\bibitem{shomshekova2023}
Shomshekova, S.; Kondratyeva, L.; Omarov, C.; Izmailova, I.; Umirbayeva, A.; Moshkina, S. Digital archival spectral data for Seyfert galaxies and their use in conjunction with modern FAI spectral data. {\em Exp. Astron.} {\bf 2023}, {\em 56}, 557--568. \url{https://doi.org/10.1007/s10686-023-09916-6}.

\bibitem{izmailova2026}
Izmailova, I.M.; Umirbayeva, A.Zh.; Khassanov, M.K.; Aktay, L.; Shomshekova, S.A. ArXSP: A python-based modular application for the reduction of digitized archival spectra. {\em Astron. Comput.} {\bf 2026}, {\em 55}, 101050. \url{https://doi.org/10.1016/j.ascom.2025.101050}.

\bibitem{umirbayeva2025}
Umirbayeva, A.Zh.; Aktay, L.; Kondratyeva, L.N.; Izmailova, I.M.; Shomshekova, S.A. Methodology for the Reduction of Archival Slit Spectra of Planetary Nebulae. {\em Acad. J. Phys. Chem. Sci.} {\bf 2025}, {\em 3}, 115--130. \url{https://doi.org/10.32014/2025.2518-1483.368} (In Russian)

\bibitem{demleitner2022plates}
Demleitner, M.; Tuvikene, T.; Schmalz, S.; Enke, H.; Partl, A. FITS Headers for Scans of Photographic Plates (Version 1.0). { IVOA Endorsed Note}. { 2022}. Available online: \url{http://www.ivoa.net/documents/Notes/PlateHeaders/} (accessed on 23~September~2026).

\bibitem{kazvo_fits_keywords}
KazVO Standard FITS Header Keywords Specification. Available online: \url{https://vo.fai.kz/kazvo-fits-keywords.html} (accessed on 23~September~2026).

\bibitem{lang2010}
Lang, D.; Hogg, D.W.; Mierle, K.; Blanton, M.; Roweis, S. Astrometry.net: Blind astrometric calibration of arbitrary astronomical images. {\em Astron. J.} {\bf 2010}, {\em 139}, 1782--1800. \url{https://doi.org/10.1088/0004-6256/139/5/1782}

\bibitem{astrometry_index_url}
Astrometry.net team. Astrometry.net Index Files and Pre-Built Index Datasets. Available online: \url{https://data.astrometry.net/} (accessed on 23~September~2026).

\bibitem{astropy:2013}
Astropy Collaboration; Robitaille, T.P.; Tollerud, E.J.; Greenfield, P.; Droettboom, M.; Bray, E.; Aldcroft, T.; Davis, M.; Ginsburg, A.; Price-Whelan, A.M.; et al. Astropy: A community Python package for astronomy. {\em Astron. Astrophys.} {\bf 2013}, {\em 558}, A33. \url{https://doi.org/10.1051/0004-6361/201322068}

\bibitem{astropy:2018}
Astropy Collaboration; Price-Whelan, A.M.; Sip{\H{o}}cz, B.M.; G{\"u}nther, H.M.; Lim, P.L.; Crawford, S.M.; Conseil, S.; Shupe, D.L.; Craig, M.W.; Dencheva, N.; et al. The Astropy Project: Building an Open-science Project and Status of the v2.0 Core Package. {\em Astron. J.} {\bf 2018}, {\em 156}, 123. \url{https://doi.org/10.3847/1538-3881/aabc4f}

\bibitem{astropy:2022}
Astropy Collaboration; Price-Whelan, A.M.; Lim, P.L.; Earl, N.; Starkman, N.; Bradley, L.; Shupe, D.L.; Patil, A.A.; Corrales, L.; Brasseur, C.E.; et al. The Astropy Project: Sustaining and Growing a Community-oriented Open-source Project and the Latest Major Release (v5.0) of the Core Package. {\em Astrophys. J.} {\bf 2022}, {\em 935}, 167. \url{https://doi.org/10.3847/1538-4357/ac7c74}

\bibitem{matt_craig_2017_1069648}
Craig, M.; Crawford, S.; Seifert, M.; Robitaille, T.; Sip{\H{o}}cz, B.; Walawender, J.; Vin{\'{\i}}cius, Z.; Ninan, J.P.; Droettboom, M.; Youn, J.; et al. astropy/ccdproc: v1.3.0.post1. {Zenodo}. {2017}. Available online: \url{https://zenodo.org/records/1069648} (accessed on 3~November~2026)

\bibitem{fai_sort_repo}
Fesenkov Astrophysical Institute. Observational Data Sorting and Ingestion Scripts. Available online: \url{https://github.com/fai-kz/obsdata_sort_analyse} (accessed on 23~September~2026).

\bibitem{1996A&AS..117..393B}
Bertin, E.; Arnouts, S. SExtractor: Software for source extraction. {\em Astron. Astrophys. Suppl. Ser.} {\bf 1996}, {\em 117}, 393--404. \url{https://doi.org/10.1051/aas:1996164}

\bibitem{Barbary2016}
Barbary, K. SEP: Source Extractor as a library. {\em J. Open Source Softw.} {\bf 2016}, {\em 1}, 58. \url{https://doi.org/10.21105/joss.00058}


\bibitem{ffdb_repo}
Fesenkov Astrophysical Institute. FFDB API Server for FAI FITS Database Infrastructure. Available online: \url{https://github.com/fai-kz/ffdb-api-server} (accessed on 23~September~2026).

\bibitem{gluchshenko2026}
Gluchshenko, A.; Izmailova, I.; Umirbayeva, A.; Kuvatova, D.; Yurin, D. Automated Wavelength Calibration of Astronomical Spectra with DTW-RANSAC Matching and CNN-Based Preprocessing. {\em Her. Kazakh-Br. Tech. Univ.} {\bf 2026}, {\em 23, 3}, 461--472. \url{https://doi.org/10.55452/1998-6688-2026-23-3-461-472} 

\bibitem{rots2022coords}
Rots, A.; Cresitello-Dittmar, M.; Laurino, O. Astronomical Coordinates, Coordinate Systems Version 1.0.
 { IVOA Recommendation} { 2022}. \url{https://ui.adsabs.harvard.edu/abs/2022ivoa.specQ1004R/abstract} (accessed on 3~November~2026).

\bibitem{rots2022meas}
Rots, A.; Cresitello-Dittmar, M. Astronomical Measurements Model Version 1.0. { IVOA Recommendation}. { 2022}. \url{https://ui.adsabs.harvard.edu/abs/2022ivoa.spec.1004R} (accessed on 23~September~2026).

\bibitem{genova2019ucd}
Genova, F.; Louys, M.; Preite Martinez, A.; Cecconi, B.; Derrière, S.; Molinaro, M.; Delmotte, N.; Gray, N.; Mann, R.; McDowell, J.; et al. Maintenance of the List of UCD Words Version 2.0. { IVOA Recommendation}. { 2019}. 
\url{https://ui.adsabs.harvard.edu/abs/2019ivoa.spec.1007G/abstract} (accessed on 3~November~2026).

\bibitem{derriere2014vounits}
Derriere, S.; Gray, N.; Demleitner, M.; Louys, M.; Ochsenbein, F. Units in the VO Version 1.0. { IVOA Recommendation}. { 2014}. Available online:
\url{https://ui.adsabs.harvard.edu/abs/2014ivoa.spec.0523D/abstract} (accessed on 3~November~2026).

\bibitem{demleitner2023vocabularies}
Demleitner, M.; Gray, N.; Taylor, M. Vocabularies in the VO Version 2.1. { IVOA Recommendation}. {2023}. \url{https://ui.adsabs.harvard.edu/abs/2023ivoa.spec.0206D} (accessed on 23~September~2026).

\bibitem{ochsenbein2019votable}
Ochsenbein, F.; Taylor, M.; Donaldson, T.; Williams, R.; Davenhall, C.; Demleitner, M.; Durand, D.; Fernique, P.; Giaretta, D.; Hanisch, R.; et al. VOTable Format Definition Version 1.4. { IVOA Recommendation} { 2019}. Available online:
\url{https://ui.adsabs.harvard.edu/abs/2019ivoa.spec.1021O/abstract} (accessed on 3~November~2026).

\bibitem{osuna2008adql}
Osuna, P.; Ortiz, I.; Lusted, J.; Dowler, P.; Szalay, A.; Shirasaki, Y.; Nieto-Santisteban, M.A.; Ohishi, M.; O'Mullane, W. IVOA Astronomical Data Query Language Version 2.00. { IVOA Recommendation}. { 2008}. Available online:
\url{https://ui.adsabs.harvard.edu/abs/2008ivoa.spec.1030O/abstract} (accessed on 3~November~2026).

\bibitem{plante2008scs}
Plante, R.; Williams, R.; Hanisch, R.; Szalay, A. Simple Cone Search Version 1.03. { IVOA Recommendation}. { 2008}. Available online:
\url{https://ui.adsabs.harvard.edu/abs/2008ivoa.specQ0222P/abstract} (accessed on 3~November~2026).

\bibitem{bonnarel2023datalink}
Bonnarel, F.; Dowler, P.; Michel, L.; Demleitner, M.; Taylor, M. IVOA DataLink Version 1.1. { IVOA Recommendation}. { 2023}. \url{https://ui.adsabs.harvard.edu/abs/2023ivoa.spec.1215B} (accessed on 23~September~2026).

\bibitem{bonnarel2017soda}
Bonnarel, F.; Dowler, P.; Demleitner, M.; Tody, D.; Dempsey, J. IVOA Server-Side Operations for Data Access Version 1.0. {IVOA Recommendation}. { 2017}. Available online:
\url{https://ui.adsabs.harvard.edu/abs/2017ivoa.spec.0517B/abstract} (accessed on 3~November~2026).

\bibitem{swinbank2017vtp}
Swinbank, J.D.; Allan, A.; Denny, R.B. VOEvent Transport Protocol Version 2.0. { IVOA Recommendation} { 2017}. Available online:
\url{https://ui.adsabs.harvard.edu/abs/2017ivoa.spec.0320S/abstract} (accessed on 3~November~2026).

\bibitem{demleitner2016identifiers}
Demleitner, M.; Plante, R.; Linde, T.; Williams, R.; Noddle, K. IVOA Identifiers Version 2.0. {IVOA Recommendation} { 2016}. Available online:
\url{https://ui.adsabs.harvard.edu/abs/2016ivoa.spec.0523D/abstract} (accessed on 3~November~2026).

\bibitem{demleitner2019regtap}
Demleitner, M.; Harrison, P.; Molinaro, M.; Greene, G.; Dower, T.; Perdikeas, M. IVOA Registry Relational Schema Version 1.1. { IVOA Recommendation}. { 2019}. Available online:
\url{https://doi.org/10.5479/ADS/bib/2019ivoa.spec.1011D} (accessed on 3~November~2026).

\bibitem{plante2018voresource}
Plante, R.; Demleitner, M.; Benson, K.; Graham, M.; Greene, G.; Harrison, P.; Lemson, G.; Linde, T.; Rixon, G. VOResource: An XML Encoding Schema for Resource Metadata Version 1.1. { IVOA Recommendation}. { 2018}. Available online:
\url{https://ui.adsabs.harvard.edu/abs/2018ivoa.spec.0625P/abstract} (accessed on 3~November~2026).

\bibitem{demleitner2012tapregext}
Demleitner, M.; Dowler, P.; Plante, R.; Rixon, G.; Taylor, M. TAPRegExt: A VOResource Schema Extension for Describing TAP Services Version 1.0. { IVOA Recommendation}. { 2012}. Available online:
\url{https://ui.adsabs.harvard.edu/abs/2012ivoa.spec.0827D/abstract} (accessed on 3~November~2026).

\bibitem{graham2017vosi}
Graham, M.; Rixon, G.; Dowler, P.; Major, B. IVOA Support Interfaces Version 1.1. { IVOA Recommendation}. { 2017}. Available online:
\url{https://ui.adsabs.harvard.edu/abs/2017ivoa.spec.0524G/abstract} (accessed on 3~November~2026).

\bibitem{taylor2017topcat}
Taylor, M. TOPCAT: Desktop Exploration of Tabular Data for Astronomy and Beyond. {\em Informatics} {\bf 2017}, {\em 4}, 18. \url{https://doi.org/10.3390/informatics4030018}

\bibitem{bonnarel2000aladin}
Bonnarel, F.; Fernique, P.; Bienaym{\'e}, O.; Egret, D.; Genova, F.; Louys, M.; Ochsenbein, F.; Wenger, M.; Bartlett, J.G. The ALADIN interactive sky atlas: A reference tool for identification of astronomical sources. {\em Astron. Astrophys. Suppl. Ser.} {\bf 2000}, {\em 143}, 33--40. \url{https://doi.org/10.1051/aas:2000331}

\bibitem{graham2014pyvo}
Graham, M.; Plante, R.; Tody, D.; Fitzpatrick, M. PyVO: Python Access to the Virtual Observatory. {\em Astrophys. Source Code Libr.} {\bf 2014}. 
\url{https://ui.adsabs.harvard.edu/abs/2014ascl.soft02004G/abstract} (accessed on 3~November~2026).

\bibitem{boch2010samp}
Boch, T.; Fitzpatrick, M.; Taylor, M.; Allan, A.; Paioro, L.; Taylor, J.; Tody, D. Simple Application Messaging Protocol Version 1.2. { IVOA Recommendation}. {2010}. Available online: \url{https://ui.adsabs.harvard.edu/abs/2010ivoa.spec.1216B/abstract} (accessed on 3~November~2026).

\bibitem{kazvo_policy}
KazVO Data Access Policy. Available online: \url{https://vo.fai.kz/kazvo-data-policy.html} (accessed on 3~November~2026).

\bibitem{eso_policy}
ESO Data Access Policy. Available online: \url{https://archive.eso.org/cms/eso-data-access-policy.html} (accessed on 23~September~2026).

\bibitem{hst_policy}
Hubble Data Policy. Available online: \url{https://science.nasa.gov/mission/hubble/observatory/science-operations/} (accessed on 23~September~2026).

\bibitem{chandra_policy}
Chandra X-Ray Center Proprietary Data Policy. Available online: \url{https://cxc.harvard.edu/cda/public.html} (accessed on 23~September~2026).

\bibitem{nrao_policy}
NRAO User Policies: Proprietary Periods and Observational Data Access. Available online: \url{https://science.nrao.edu/observing/policies/docs/manuals/users-policy/preparation-and-execution-of-observations/data-delivery-and-data-rights/proprietary-periods-and-observational-data-access} (accessed on 23~September~2026).

\bibitem{keck}
W. M. Keck Observatory Data Access and Proprietary Policy. Available online: \url{https://koa.ipac.caltech.edu/UserGuide/proprietary_policy.html} (accessed on 23~September~2026).

\bibitem{iso8601}
\emph{ISO 8601-1:2019}; Date and Time---Representations for Information Interchange---Part 1: Basic Rules. International Organization for Standardization: Geneva, Switzerland, 2019.

\bibitem{2019ApJS..241...33L}
Li, Y.-R.; Wang, J.-M.; Zhang, Z.-X.; Wang, K.; Huang, Y.-K.; Lu, K.-X.; Hu, C.; Du, P.; Bon, E.; Ho, L.C.; et al. A Possible $\sim$20 yr Periodicity in Long-term Optical Photometric and Spectral Variations of the Nearby Radio-quiet Active Galactic Nucleus Ark 120. \textit{Astrophys. J. Suppl. Ser.} \textbf{2019}, \textit{241}, 33. \url{https://doi.org/10.3847/1538-4365/ab0ec5} (accessed on 3~November~2026).

\bibitem{Shomshekova2025}
Shomshekova, S.; Reva, I.; Kondratyeva, L.; Huseynov, N.; Kim, V.; Aktay, L. Spectral and Photometric Studies of NGC 7469 in the Optical Range. \textit{Universe} \textbf{2025}, \textit{11}, 227. \url{https://doi.org/10.3390/universe11070227} (accessed on 3~November~2026).

\bibitem{fruscione2026ds9}
Fruscione, A.; Glotfelty, K.; Joye, W.; McDowell, J. SAOImageDS9: An Essential Tool for Astronomical Exploration. {\em Astron. J.} {\bf 2026}, {\em 172}, 116. \url{https://doi.org/10.3847/1538-3881/ae84c1} (accessed on 3~November~2026).

\end{thebibliography}
\end{document}